\documentclass[twocolumn,tighten]{aastex62} 
\usepackage{graphicx}
\usepackage{color}
\DeclareGraphicsExtensions{.eps,.png,.jpg}

\def\aj{{AJ}}

\def\apj{{ApJ}}
\def\apjs{{ApJS}}
\def\asec{$^{\prime\prime}$}

\def\kms{km~s$^{-1}$}

\def\lax{{$\mathrel{\hbox{\rlap{\hbox{\lower4pt\hbox{$\sim$}}}\hbox{$<$}}}$}}
\def\gax{{$\mathrel{\hbox{\rlap{\hbox{\lower4pt\hbox{$\sim$}}}\hbox{$>$}}}$}}
\def\simlt{\lower.5ex\hbox{$\; \buildrel < \over \sim \;$}}
\def\simgt{\lower.5ex\hbox{$\; \buildrel > \over \sim \;$}}

\def\mnras{{MNRAS}}

\def\sb{mag~arcsec$^{-2}$}

\begin{document}
\title{Spectroscopic Constraints on the Build-up of the Intracluster Light in the Coma Cluster}

\author[0000-0002-4267-9344]{Meng Gu}
\affiliation{Department of Astronomy, Harvard University, Cambridge, MA 02138, USA}
\author[0000-0002-1590-8551]{Charlie Conroy}
\affiliation{Department of Astronomy, Harvard University, Cambridge, MA 02138, USA}
\author[0000-0002-9402-186X]{David Law}
\affiliation{Space Telescope Science Institute, 3700 San Martin Drive, Baltimore, MD 21218, USA}
\author[0000-0002-8282-9888]{Pieter van Dokkum}
\affiliation{Astronomy Department, Yale University, New Haven, CT 06511, USA}
\author[0000-0003-1025-1711]{Renbin Yan}
\affiliation{Department of Physics and Astronomy, University of Kentucky, 505 Rose Street, Lexington, KY 40506-0057, USA}
\author[0000-0002-6047-1010]{David Wake}
\affiliation{Department of Physical Sciences, The Open University, Milton Keynes, MK7 6AA, UK}
\author[0000-0001-9742-3138]{Kevin Bundy}
\affiliation{Department of Astronomy and Astrophysics, University of California, Santa Cruz, CA 95064, USA}
\author[0000-0003-1887-0621]{Alexa Villaume}
\affiliation{Department of Astronomy and Astrophysics, University of California, Santa Cruz, CA 95064, USA}
\author[0000-0002-4542-921X]{Roberto Abraham}
\affiliation{Department of Astronomy and Astrophysics, University of Toronto, 50 St. George Street, Toronto, ON M5S 3H4, Canada}
\author[0000-0001-9467-7298]{Allison Merritt}
\affiliation{Max-Planck-Institut f\"ur Astronomie, K\"onigstuhl 17, D-69117 Heidelberg, Germany}
\author[0000-0001-5310-4186]{Jielai Zhang}
\affiliation{Schmidt Science Fellows, in Partnership with the Rhodes Trust, Rhodes House, Oxford, OX1 3RG, UK}
\author[0000-0002-3131-4374]{Matthew Bershady}
\affiliation{Department of Astronomy, University of Wisconsin-Madison, 475N. Charter Street, Madison WI 53703, USA}
\affiliation{South African Astronomical Observatory, P.O. Box 9, Observatory 7935, Cape Town, South Africa}
\author[0000-0002-3601-133X]{Dmitry Bizyaev}
\affiliation{Apache Point Observatory, P.O. Box 59, Sunspot, NM 88349, USA}
\affiliation{Sternberg Astronomical Institute, Moscow State University, Moscow, Russia}
\author[0000-0002-2835-2556]{Kaike Pan}
\affiliation{Apache Point Observatory, P.O. Box 59, Sunspot, NM 88349, USA}
\author[0000-0002-6325-5671]{Daniel Thomas}
\affiliation{Institute of Cosmology \& Gravitation, University of Portsmouth, Dennis Sciama Building, Portsmouth, PO1 3FX, UK}
\author[0000-0002-5908-6852]{Anne-Marie Weijmans}
\affiliation{School of Physics and Astronomy, University of St. Andrews, North Haugh, St. Andrews KY16 9SS, UK}
\begin{abstract}

The stellar content of the intracluster light (ICL) provides unique insight 
into the hierarchical assembly process of galaxy clusters.  However, the ICL 
is difficult to study due to its low surface brightness and large physical extent. 
We present optical spectra of three ICL regions in the Coma cluster, 
located between 100-180~kpc from their nearest brightest cluster galaxies 
(BCGs): NGC~4889 and NGC~4874.  The mean surface brightness of the three 
ICL regions are $\mu_g\approx25.3-26.2$~\sb.  Integral-field unit (IFU) 
spectroscopy with 13.5 hr on-source integration time were acquired as part 
of an ancillary program within the SDSS-IV MaNGA survey.  We stacked the 
127 individual fiber spectra in each IFU in order to achieve a $1\sigma$ 
limiting surface brightness of 27.9~\sb, corresponding to a mean 
signal-to-noise ratio in the optical of 21.6\AA$^{-1}$, 9.6\AA$^{-1}$, 
and 11.6\AA$^{-1}$, for each region.  We apply stellar population models to the stacked
spectra, and measure the recession velocities, velocity dispersions, 
stellar ages, and metallicities in the three ICL regions.  Our results 
show that the velocity dispersions of ICL regions are very high 
($\sigma\sim 650$~\kms), indicating the stars in these regions are 
tracing the gravitational potential of the cluster, instead of any 
individual galaxy.  The line-of-sight velocities of the three ICL 
regions are different from each other by $\sim700$~\kms, while the 
velocity of each region is similar to the closest BCG.  This suggests 
that the ICL regions are associated with two distinct subclusters 
centered on NGC~4889 and NGC~4874.
The stellar populations of these 
regions are old and metal poor, with ages of 
$12.4^{+1.2}_{-3.3}$~Gyr, $6.6^{+2.5}_{-2.4}$~Gyr, and $10.6^{+2.8}_{-4.3}$~Gyr, 
and iron abundances, [Fe/H], of  $-0.8^{+0.2}_{-0.3}$, 
$-0.6^{+0.3}_{-0.5}$, and $-0.8^{+0.3}_{-0.3}$.  From the derived age 
and metallicity, the build-up of ICL in Coma is likely to be through the 
accretion of low mass galaxies or the tidal stripping of the outskirts 
of massive galaxies that have ended their star formation early on, 
instead of directly from major mergers of massive galaxies.  

\end{abstract}
\keywords{galaxies: evolution}
\section{Introduction}

According to the widely accepted $\Lambda$-Cold Dark Matter model, massive 
early type galaxies (ETGs) are assembled hierarchically following 
their underlying dark matter structures \citep{White1978}. The evolution of ETGs 
in massive halos can be described by a two-phase picture \citep{Naab2007,Feldmann2010,Johansson2012,Navarro-Gonzalez2013,
Oser2010,Oser2012,Qu2017,Rodriguez-Gomez2016,Lackner2012}: 
at high redshift their evolution is dominated by the concentrated mass growth 
through rapid dissipational in-situ star formation 
\citep{Daddi2005,Trujillo2006,Hopkins2008,Dekel2009,Hyde2009} 
At later times, their evolution is increasingly dominated by the build up of 
the outskirts through multiple mergers and accretions of lower mass galaxies 
\citep{Ostriker1975, vanderWel2014,Bezanson2009,vanDokkum2010,vanderWel2011}.  

Brightest cluster galaxies (BCGs) are a special class of ETGs residing near 
the center of galaxy clusters. The most significant structural feature of BCGs 
is their diffuse and extended stellar envelopes, so they are also classified 
as cD galaxies.  If we trace the 
stellar distribution from the inner regions of BCGs to the stellar envelopes,  
part of the stellar components would be no longer bound to the galaxy, but 
instead associated with the whole cluster as we approach large radius.  
Many studies have confirmed that in some massive ETGs and BCGs, the stellar 
velocity dispersion profiles rise with increasing radii towards the velocity 
dispersion of the cluster \citep[e.g.][]{Faber1977,Dressler1979,Kelson2002,
Bender2015,Veale2018}.  The stellar structure surrounding BCGs that are  
gravitationally bounded to the galaxy cluster is called the intracluster 
light (ICL) \citep{Zwicky1951}.  The formation of the ICL is considered to be a 
combined effect of multiple mechanisms, including tidal disruption of dwarf galaxies 
\citep{Rudick2009}, tidal striping of low mass galaxies through galaxy 
interactions \citep{Conroy2007, Purcell2007, Rudick2009, Contini2014}, 
violent relaxation during major mergers \citep{Murante2007}, 
and in-situ star formation \citep{Puchwein2010}.  Stellar populations in 
the outskirts of galaxies and the ICL, if measured, can help constrain these scenarios.

\begin{figure*}[t] 
\centering 
\includegraphics[width=14.7cm]{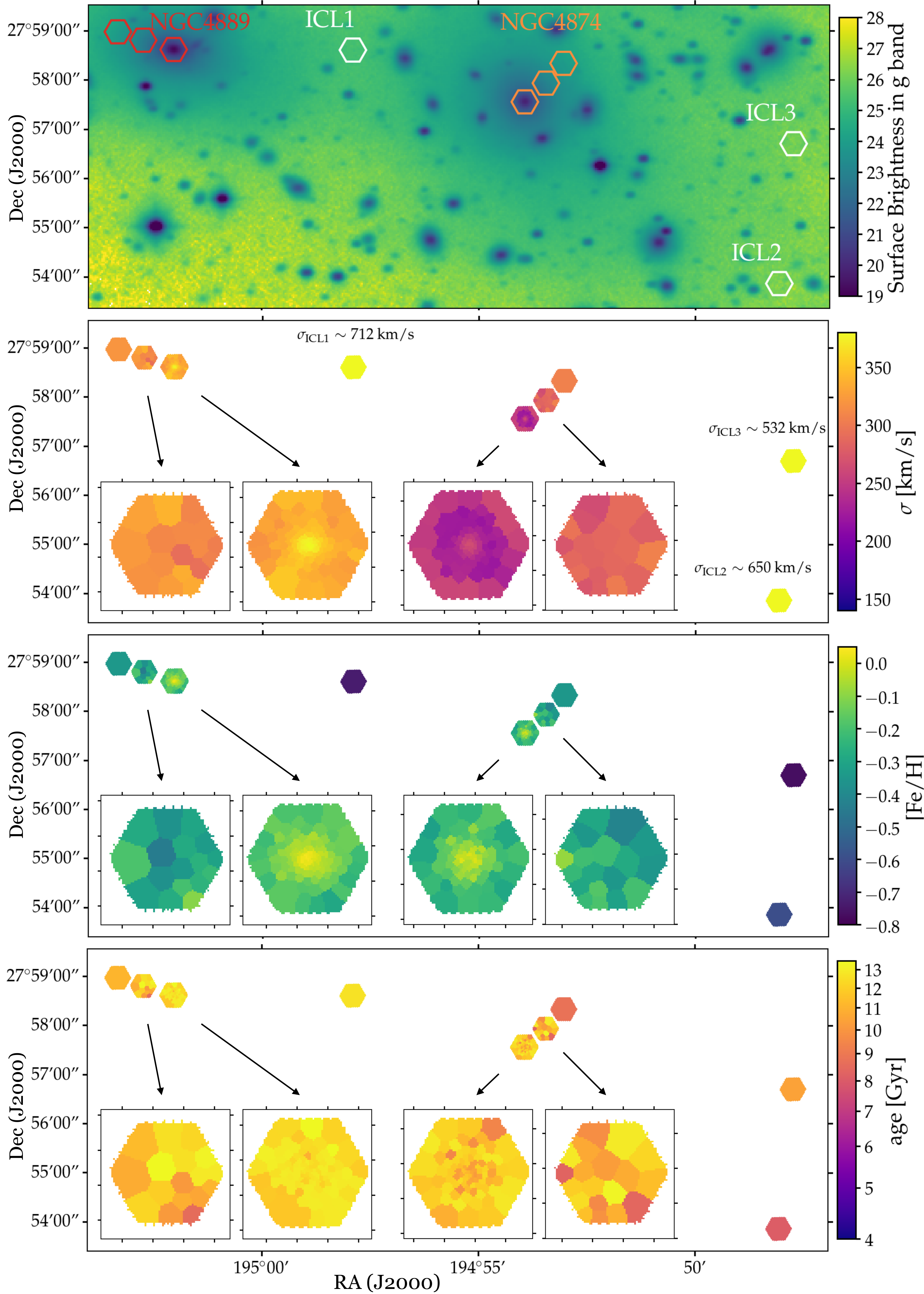}
\caption{
 First panel: overview of the IFU bundle locations on g-band surface brightness map 
 observed by the Dragonfly Telephoto Array.  Red hexagons show IFU locations on 
 the center, $1R_{\rm e}$ and $2R_{\rm e}$ of NGC~4889.  Orange hexagons show IFU locations on the 
 center, $0.5R_{\rm e}$, $1R_{\rm e}$ of NGC~4874.  White hexagons show locations of IFUs on ICL1, 
 ICL2, and ICL3, respectively. Second to last panels: spatial distributions of 
 best-fit parameters derived from fitting spatially stacked spectra by {\tt alf}: 
 stellar velocity dispersion $\sigma$, iron abundance [Fe/H] and stellar age.   
 }
\label{figure1}
\end{figure*}
\noindent

Observing the outskirts of massive ETGs is difficult due to the 
low surface brightness, but this field has been accelerated by the 
state of art instruments and improved data reduction in  
both photometry 
\citep{vanDokkum2014,Duc2015,Mihos2005,Huang2016,Huang2018a,Huang2018b} 
and spectroscopy 
\citep{Sanchez-Blazquez2007,Foster2009,Spolaor2010,Greene2012,Greene2015,Coccato2010,Coccato2011}.  
Observations of the diffuse ICL with extremely low surface brightness is 
even more challenging.  Deep imaging and accurate sky subtraction are required 
to detect the ICL and constrain their growth over time \citep{Burke2015,Gonzalez2005,Rudick2010,Toledo2011,Guennou2012,Giallongo2014, Mihos2005, 
Zibetti2005}. 
The stellar population and kinematic properties have been explored through 
multi wavelength photometry \citep{Montes2014,Williams2007,Montes2018}, 
integral-field spectroscopy \citep{Adami2016, Edwards2016}, globular clusters 
\citep{AlamoMartinez2017}, and individual objects such as planetary nebulae 
\citep{Arnaboldi2003, Arnaboldi2004, Gerhard2007} and red giants branch stars 
\citep{Feldmeier2004, Longobardi2015}.

In this paper, we present a stellar population analysis from the 
BCG centers out to the ICL regime through full spectral modeling, 
and provide the recession velocity, velocity dispersion, stellar age, 
and iron abundance out to 180~kpc.  This is the first time 
that the stellar population analysis through full optical spectra 
modeling is performed beyond 100~kpc.
We make use of the data obtained as part of the Deep Coma ancillary 
program within the SDSS-IV/MaNGA program, and the $g$ and $r$ band 
photometry from the Dragonfly Telephoto Array \citep{Abraham2014}.  
The Coma cluster has a median redshift of $cz=7090$~\kms 
\citep{Geller1999} and a velocity dispersion of $\sigma\sim1000$~\kms 
\citep{Colless1996, Mobasher2001, Rines2013, Sohn2016}. 
The distance of Coma Cluster is assumed to be 100.0 Mpc, 
adopted from Liu \& Graham (2001). This corresponds to a distance 
modulus of 34.99 mag and a scale of 0.474~kpc~arcsec$^{-1}$.  
The Galactic foreground extinction for Coma Cluster is $A_g=0.030$ mag and 
$A_r=0.021$ mag (Schlafly \& Finkbeiner 2011).

\section{Data}

\subsection{Project Overview and Observation Strategy} 
We make use of data obtained by the MaNGA Survey (Mapping Nearby Galaxies 
at Apache Point Observatory, \citep{Bundy2015, Yan2016b, Drory2015, Wake2017,Abolfathi2017}.  
MaNGA is a large, optical integral field spectroscopy survey with 17 
 deployable integral field units (IFUs) (ranging from 12\asec 
 to 32\asec in diameter ), and one of the fourth-generation Sloan Digital 
 Sky Survey (SDSS-IV) programs 
 \citep{Gunn2006,Blanton2017}.  The primary goal of MaNGA is to obtain  integral field 
 spectroscopy of $\sim10,000$ nearby galaxies.  

MaNGA utilizes integral field units (IFUs) from two dual-beam Baryonic 
Oscillation Spectroscopic Survey (BOSS) spectrographs (\cite{Drory2015}) 
that are on the SDSS 2.5 meter telescope. The spectrographs have 1423 fibers 
in total that are bundled into different size of IFUs. The diameter of each 
fiber is 1.98\asec~on the sky. The wavelength coverage of the spectrographs 
is $3622-10354$\AA~with a $\sim 400$\AA~overlap from 
$\sim5900$ to $\sim6300$\AA~between the blue and red 
cameras.  The spectral resolution is 
$R=1560-2650$. 

Our data comes from one of the MaNGA's ancillary programs, the Coma Deep 
program 
\footnote{www.sdss.org/dr14/manga/manga-target-selection/ancillary-targets/coma}
\citep[also see][]{Gu2018a}.  This is the deepest MaNGA
ancillary program, consisting of six plates designed to observe specially selected targets in the Coma cluster.
The goal of the Coma Deep program is to study the stellar population 
of various kinds of targets in the Coma cluster and its surrounding area.  
The plates are $\sim0.7$~meter in diameter and $3^{\circ}$~on the sky.  
The center of all plates is at RA$=12^h 58^m 35.58^s$, DEC$=27^d36^m12.744^s$.  
Five massive ETGs are selected for observations: NGC~4889, NGC~4874, NGC~4860, 
NGC~4839 and NGC~4841A, as well as three dwarf elliptical galaxies: 
GMP~2232, GMP~5076 and GMP~5361.  Observations of massive ETGs are 
conducted on their central regions and outskirts up to $\sim40$~kpc away 
from the centers. These positions are carefully 
chosen for optimizing IFU bundle mapping of desired targets. Three 
127-fiber bundles are placed in regions of extremely low surface brightness 
in order to probe the stellar populations of the ICL.  These regions are 
selected based on deep images by the Dragonfly Telephoto Array 
and are away from any foreground contamination, with a surface brightness 
in $g$ band from 25.3 to 26.2~\sb.  The diameter of each 127-fiber 
IFU bundle on the sky is 32\asec.5.  

Locations of IFU bundles on two BCGs and three ICL regions are shown 
in the Dragonfly-$g$ band surface brightness map in Figure~1. 
The distance from ICL1 to NGC~4889 and NGC~4874 are 219\asec.2 and 
219\asec.6, respectively.  The distance from ICL2 and ICL3 to 
NGC~4874 are 382\asec.1 and 332\asec.7, respectively.

\begin{figure}[t]
\centering
\includegraphics[width=8.5cm]{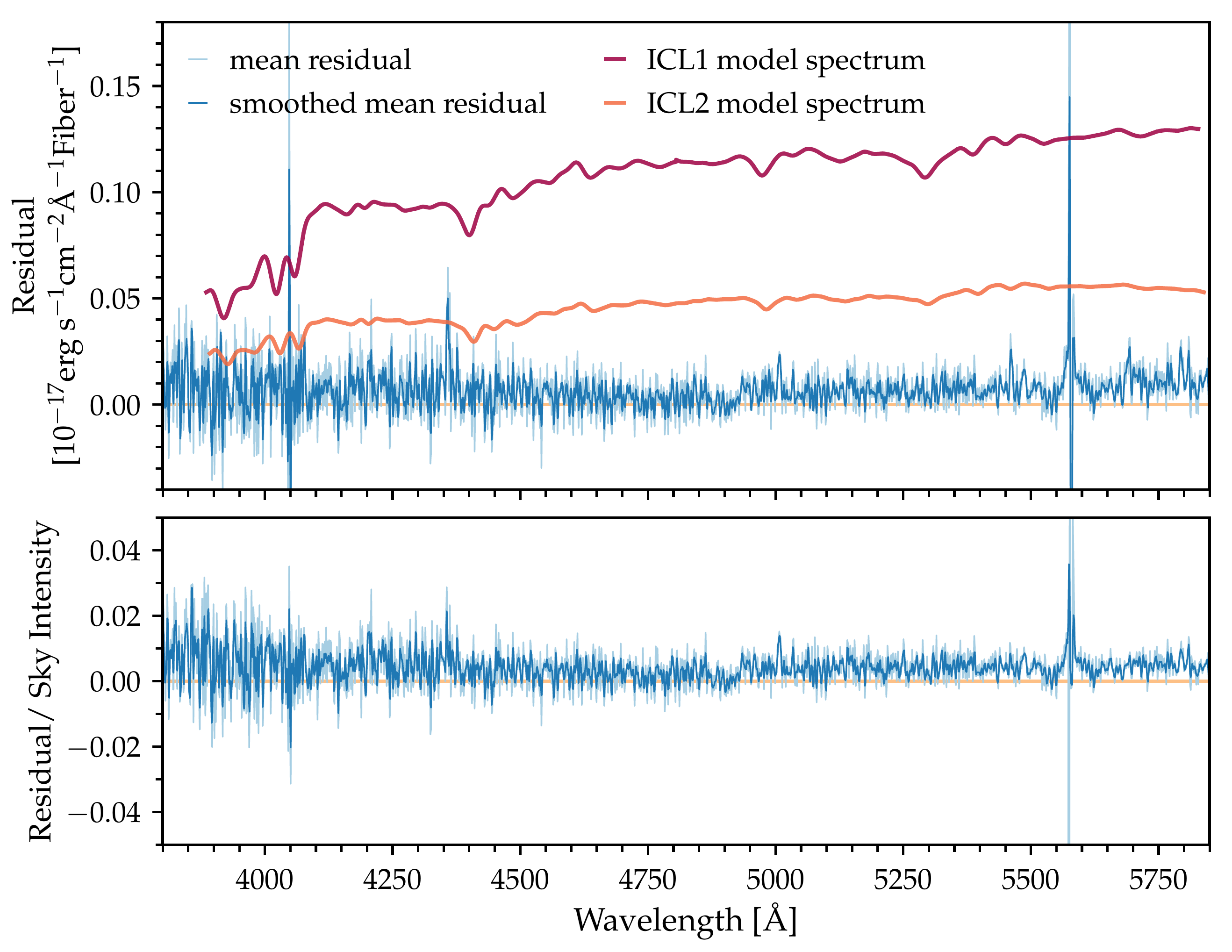}
\caption{
Top panel: mean stacked residual of sky-subtracted sky spectra from all 
science fibers in the nod exposures in which all bundles are placed on 
background sky (light blue), and that smoothed by a 3-pixel moving box 
(dark blue).  Model spectra of the brightest (ICL1, red) and fainest 
(ICL2, orange) ICL regions are shown as a noiseless version of the 
expected target flux level.  
Bottom panel:  fractional residual relative to the sky intensity.  
}
\label{figure2}
\end{figure}

Two 127-fiber IFUs are used to observe NGC~4889 and NGC~4874.  We observed their 
central regions by the 1st and 2nd Coma plates.  We observed $1R_{\rm e}$ of NGC~4889 
and $0.5R_{\rm e}$ of NGC~4874 with the third and fourth Coma plates.  They are 
located 39\asec.0 
and 34\asec.5 to the centers, respectively.  
IFU bundles are placed at $2R_{\rm e}$ of NGC~4889 and $1R_{\rm e}$ of 
NGC~4874 in the 5th and 6th plates.  The locations to the centers are 71\asec.5, 66\asec.7, respectively.  
The regions are chosen to best avoid contamination from nearby sources.
A similar strategy is adopted for NGC~4860, NGC4839 and NGC~4841A.  They are 
observed by 91-fiber and 61-fiber bundles.  
Three dwarf elliptical galaxies, GMP~2232, GMP~5076, and GMP~5361 are observed 
by 37-fiber bundles.  The locations of IFU bundles on ICL and dwarf elliptcal 
galaxies are kept the same throughout the six plates, providing the 
deepest MaNGA observations on single targets.  

Since the dark-time sky background at APO is $\sim22$~\sb~in $g$-band, 
excellent sky subtraction is required to probe low surface brightness regions.  
In Deep Coma plates, the locations of reference sky fibers are carefully 
selected using broad band images taken by the Dragonfly Telephoto Array, 
reaching a $g$-band surface brightness $\mu_g>27.8$~\sb~ for sky fiber 
locations ($\sim3$\asec around each fiber) based on the Dragonfly imaging. 
In addition to the 92 single fibers used to construct the model sky spectrum 
for ordinary MaNGA plates, three IFU bundles (two 19-fiber bundles, 
one 37-fiber bundle) were devoted to additional measurements of the sky, 
hence there are 167 sky fibers in total used across both spectragraphs. 
 
In addition, we adopt an on-and-off nodding strategy to improve the 
accuracy of the background estimate and to mitigate systematics. By shifting 
the whole field approximately 20$\prime$ away, we obtain reference 
``all-sky'' exposures, during which a large fraction of the sky fibers and 
science IFUs sample the blank sky.  Each of the first two plates includes 
nine 5-minute nodded sky exposures at nine different locations between the 
normal science exposures. After the first two plates, we decided to change 
the exposure time to the same as the science exposure in order to better 
constrain the systematics.  Therefore, each of the 
last four plates includes four 15-minute nodded sky exposures at four 
different locations. 

\subsection{Data Reduction}
We processed the data using a custom modified version of the MaNGA Data Reduction 
Pipeline MPL-7 \citep[DRP;][]{Law2015, Law2016}.  DRP MPL-7 will be made publicly 
available in DR15 that is planned for December 2018. 
The baseline DRP first removes detector overscan regions and quadrant-dependent 
bias and extracts the spectrum of each fiber using an optimal profile-fitting 
technique. It uses the sky fibers to create a super-sampled model for the 
background sky spectrum and subtracts this model spectrum from each of the 
science fibers.  Flux calibration is then performed on individual exposures using 
12 7-fiber IFUs targeting spectrophotometric standard stars 
\citep{Yan2016,Yan2016b}.  
Fiber spectra from the blue and red cameras are then combined together onto a 
common logarithmic wavelength solution using a cubic b-spline fit. 
These ``mgCFrame" files represent spectra of all 1423 MaNGA fibers from a 
single exposure in a row-stacked format, where each row corresponds to an 
individual 1-dimensional fiber spectrum.  The logarithmic wavelength grid runs 
from $\log\lambda$(\AA)$=3.5589$ to $\log\lambda$(\AA)$=4.0151$, 
which corresponds to 4563 spectral elements from 3621.5960 to 10353.805\AA.
In this paper we only use the data taken by the blue spectrograph.  
This allows us to avoid additional issues associated with the numerous 
bright atmospheric OH features in the red.  

As described in \citet{Gu2018a}, our analysis is possible only with 
exquisite control of detector and instrumental systematics, and 
therefore some changes to 
the DRP have been made specifically for the Deep Coma program.  
Analysis of our nodded all-sky observations 
showed evidence for low-level systematics in the detector electronics.  
Therefore we added a step to measure and remove a 0.5 e$^-$/pixel 
offset in bias between the light-sensitive detector pixels and the 
overscan region, compensating at the same time for a 
seasonally-dependent 0.1 e$^-$/pixel drift in the difference.
In addition, the amplifier-dependent gain values tended to drift 
from one exposure to the next away from nominal at the $\sim0.1\%$ level; 
we added procedures to measure and 
correct for this effect empirically using the sky fibers in each 
exposure.  Finally, we modified the DRP to be able to apply the 
flux calibration vector from the nearest ordinary science 
exposure in time to the nod exposures (for which there are no calibration 
stars in the 7-fiber mini bundles). 

\begin{figure}[t]
\centering
\includegraphics[width=8.5cm]{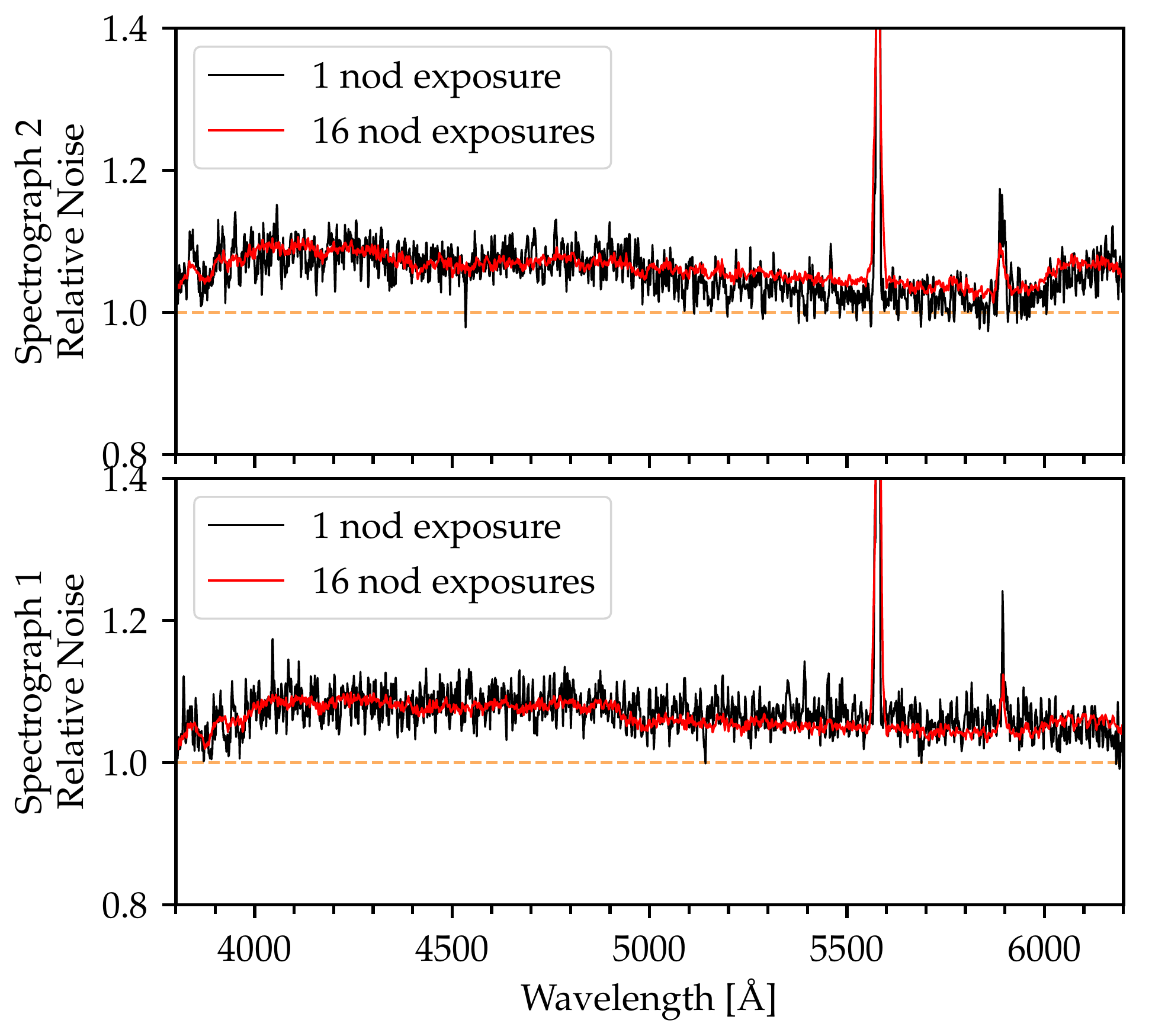}
\caption{
 Ratio between the actual noise and the expected noise from the 
 detector read noise and Poisson counting statistics as a function of 
 wavelength for the science spectra in one sky subtracted nod exposure 
 (black), and that in all 16 nod exposures (red).  The top and bottom panels 
 show results in the first and second spectrograph, respectively.  A relative noise of 
 one means perfect, ``Poisson limited'' sky subtraction (dotted line). 
}
\label{figure3}
\end{figure}
\begin{figure}[t]
\centering
\includegraphics[width=8.5cm]{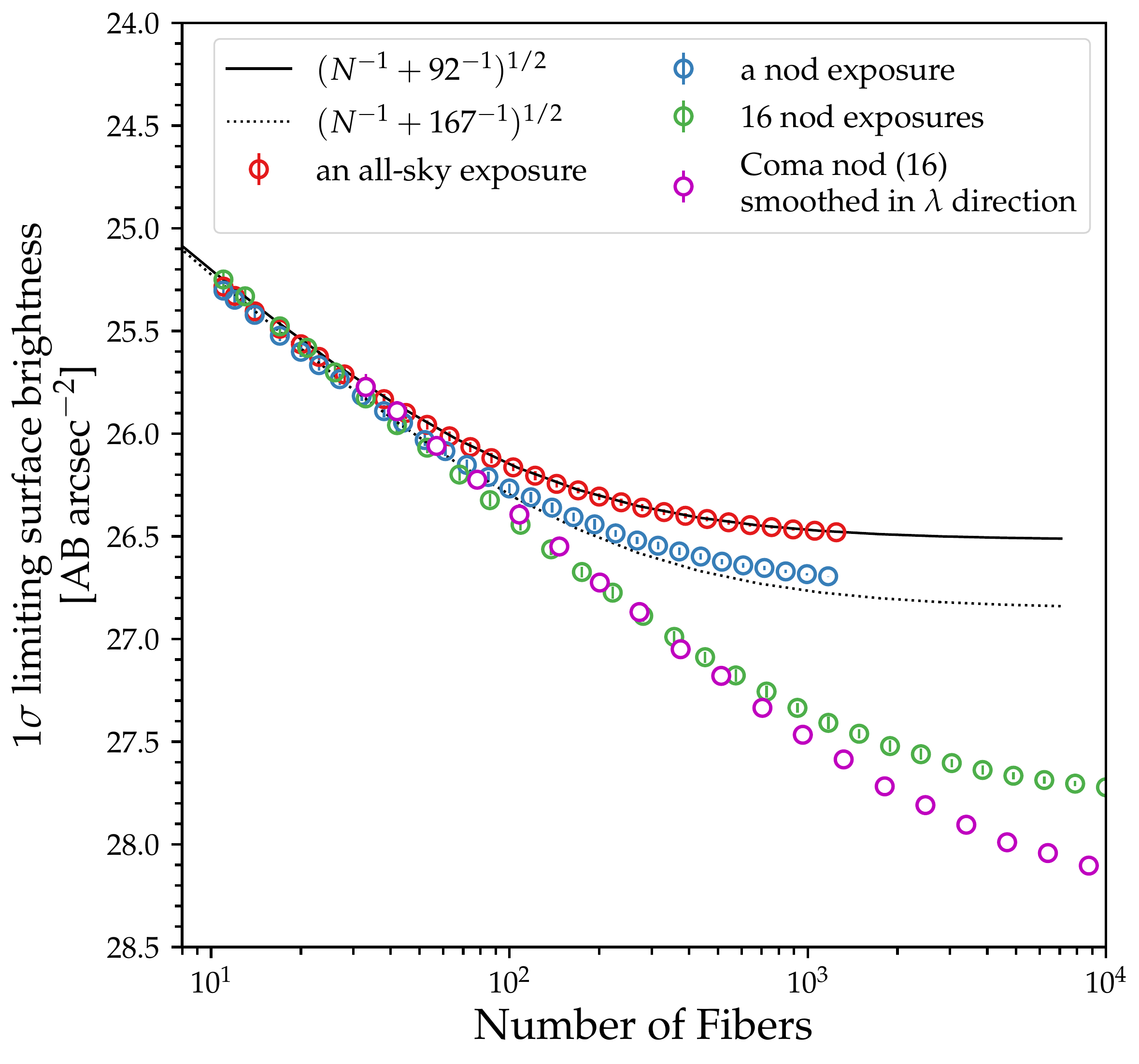}
\caption{
 $1\sigma$ limiting surface brightness in the wavelength range 
 $4000-5500$\AA~ as a function of the number of randomly selected 
 and combined fibers in a standard MaNGA all-sky exposure 
 (92 sky fibers, red), a Deep Coma nod exposure 
 (167 sky fibers, blue) and 16 nod exposures (green).  
 The solid and dotted black 
 line shows the theoretical expectation based on 
 $\sqrt{N^{-1}+92^{-1}}$ for an all-sky exposure, and 
 $\sqrt{N^{-1}+167^{-1}}$ for a nod exposure, respectively. 
 The performance of stacking across all 
 54 science exposures (purple) is estimated by smoothing the 
 stacked spectrum of all 16 nod exposures by a factor of 3 in 
 the wavelength direction.  The sky level in all exposures 
 is $\sim 22$~\sb.
}
\label{figure4}
\end{figure}
\begin{figure*}[t]
\centering
\includegraphics[width=17cm]{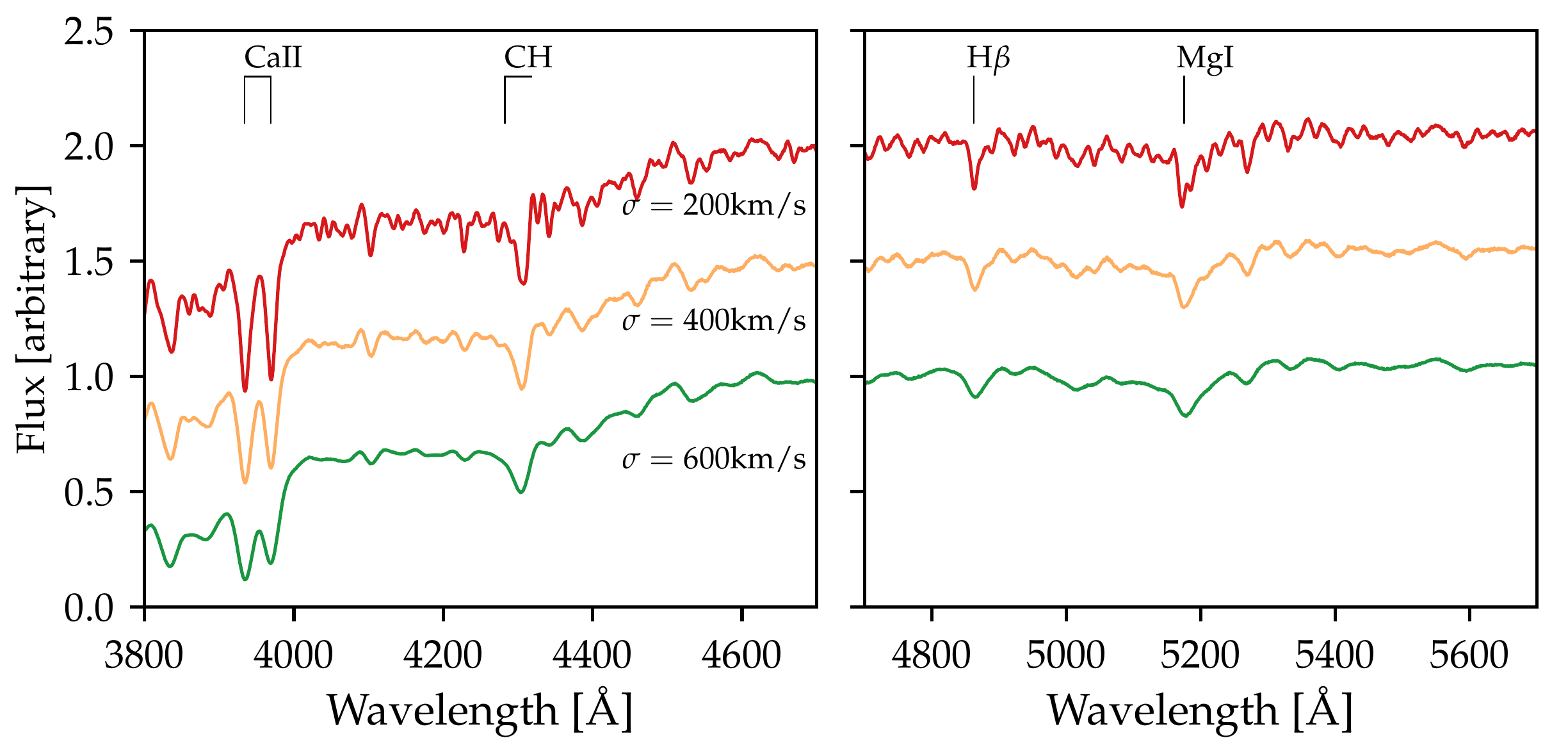}
\caption{
 Model spectra with different velocity dispersions. Targets 
 with high velocity have much shallower absorption features.  
 Model spectra for a stellar population of age of 10~Gyr and [Fe/H]$=-0.7$.  
 The spectra in red, orange and green are smoothed with velocity dispersion 
 of 200, 400 and 600~\kms.  Prominent absorption features in the ICL spectra are labeled.
}
\label{figure5}
\end{figure*}

In addition, performance analysis of early observations in the Deep
Coma program revealed that scattered light  and the extended 
($>100$ pixel) profile wings of bright galaxies targeted by the Coma 
program were contaminating the spectra of fainter objects.  We therefore 
redesigned our observing program to consolidate all bright targets 
(ETGs and dwarf ellipticals) onto one of the two BOSS spectrographs 
\citep{Smee2013}, and 
all faint targets (UDGs and ICL) onto the other so that 
these targets never share a detector.  

Although these modifications substantially improve performance for the 
Deep Coma program relative to the standard DRP, we find that the
final stacked science spectra are nonetheless still limited
by systematic residuals over wavelength scale $>100$\AA.
These residuals are consistent between stacked science and nodded 
sky spectra within each plate, possibly due to cartridge-dependent 
uncertainties in fiber alignment and the detector point-spread function.  
The offset ranges from 0 to 10$^{-18}$erg~s$^{-1}$cm$^{-2}$\AA$^{-1}$.
In the last four plates, we mitigate 
the impact of these systematics by fitting the stacked spectra of the 
sky subtracted nodded sky exposures with a 3 degree polynomial from 
3836\AA~to 5873\AA~in the observed frame and subtracting 
the result from the corresponding science exposures prior to stacking 
science spectra.   The amplitude of the polynomial correction ranges 
from $10^{-19}$ to $10^{-18}$erg~s$^{-1}$cm$^{-2}$\AA$^{-1}$ in the 
continuum, and represents an important correction to the baseline 
flux level for extremely faint targets.
We match the continuum levels of stacked spectra from 
different plates via subtracting the above polynomial continuum 
before we derive any science result.  

\subsection{Sky Subtraction Performance}

We make use of the nodded sky exposures to test the sky subtraction 
performance we can achieve for the spatially stack ICL spectra.  
We perform the same analysis on the 15-minute nodded sky exposures in 
the last four Deep Coma plates, and spatially stack the sky-subtracted 
sky spectral from all science fibers in these 16 exposures. 
This provides us an estimate of the sky residual.  
Figure~2 shows the sky subtracted residual (top panel), and the fractional 
residual relative to the sky intensity (bottom panel) from the stacked 
sky-subtracted spectra in nodded sky exposures. 
Since the number of Deep Coma science exposures is roughly 3$\times$ the 
number of nodded sky exposures, the residuals 
are smoothed by a 3-pixel moving box to mimic the residual we 
could achieve by stacking the same number of nodded sky exposures as 
the science exposures. They are shown as the dark blue lines.  
The residuals are very close to zero in the $3800-5800$\AA~: the wavelength 
range we used to perform stellar 
population analysis.  The fractional residual relative to the sky intensity 
is on average within $1\%$.  We further compare the flux level of our science 
targets, by including model spectra of ICL1 and ICL2 as a noiseless version 
of the expected flux level.  ICL1 and ICL2 are the brightest and faintest ICL 
regions, respectively.  Figure~2 shows that ICL1 has a flux level much 
higher than the sky residuals, about $9\times$ the mean sky residual in the 
$3800-5800$\AA~.  ICL2 is in general above the mean stacked residual ($4\times$) 
except for the bright sky lines, which are masked out during our spectral 
fitting procedure.   

We also examine if the sky subtraction is ``Poisson limited'' by constructing 
``Poisson ratio'' images following the procedures described in \citet{Law2016}.  
The inverse variance in the ``mgSFrame'' spectra represents the combined effect 
of shot noise and detector read noise.  By comparing the distribution of 
sky subtracted residual with the expected noise from the detector read noise 
and Poisson counting statistics, we are able to evaluate the sky subtraction 
performance in single and multiple exposures.   As shown in Figure~3, the 
distribution of Poisson ratio is slightly above 1.0 but on average smaller 
than 1.1 at all wavelength, except for the few strongest sky lines. 
Stacking across multiple exposures does not increase this ratio.  

\begin{figure*}[t]
\centering
\includegraphics[width=17cm]{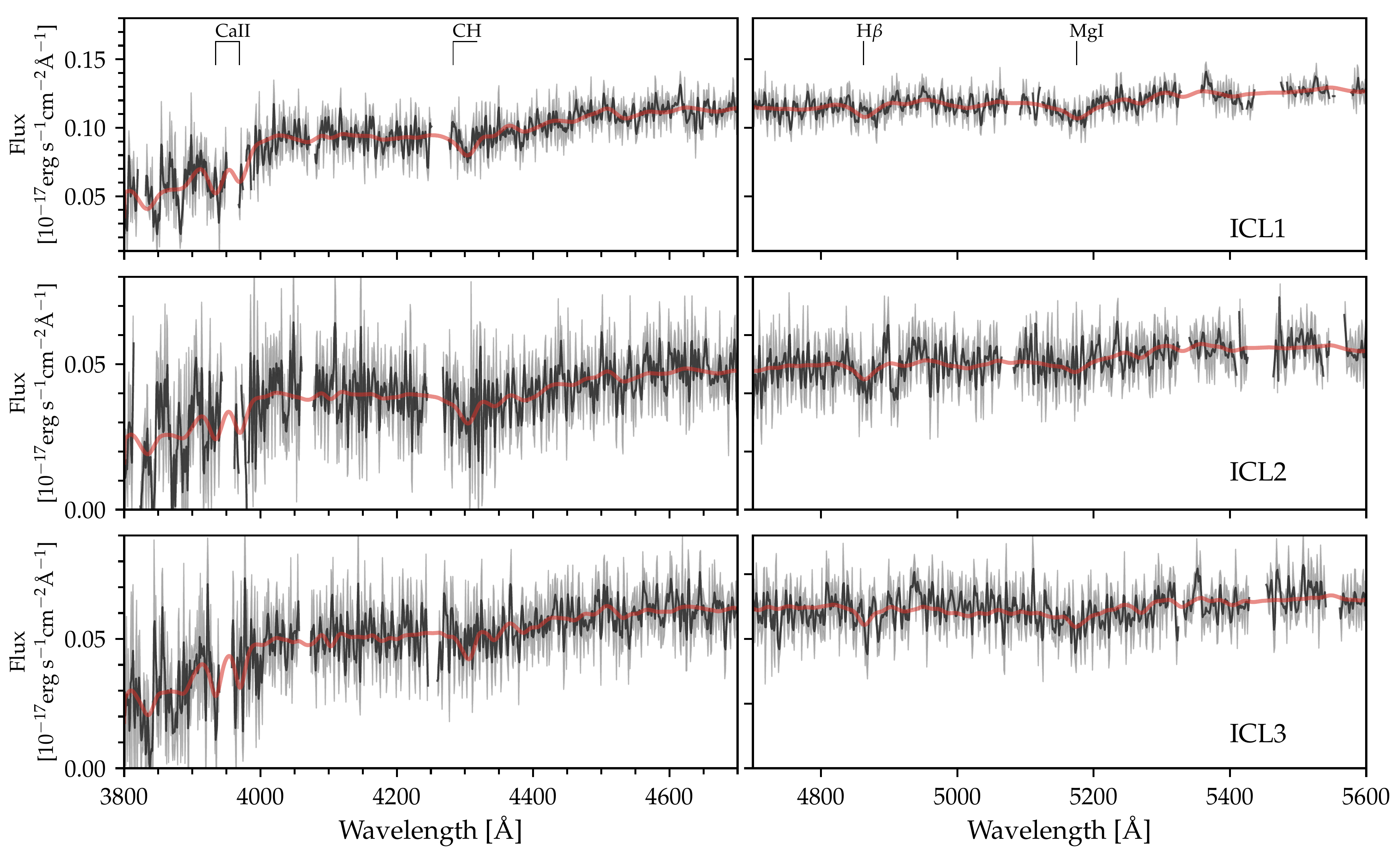}
\caption{
Stacked spectra (black) and best-fit model spectra (red) with parameters at minimum 
$\chi^2$ from {\tt alf} of ICL1 (top), ICL2 (middle), and ICL3 (bottom).  
Spectra are smoothed with a 3 pixel box car filter for the purpose of better 
display.
Gray shaded regions show the uncertainty of flux from the input spectra.  
Gaps in the black lines indicate pixels that are masked prior to the fitting, 
which are pixels contaminated by bright sky lines.}
\label{figure6}
\end{figure*}
\noindent

Furthermore, we calculate the $1\sigma$ limiting surface brightness following the 
steps in \citet{Law2016}.  Using the flux calibrated, camera combined mgCFrame 
spectra, we calculate the limiting $1\sigma$ surface brightness in 4000-5500\AA~ 
achieved in the blank sky by randomly stacking different numbers of spectra.  
The results are shown in Figure~4.  For a typical MaNGA all-sky exposure with 
92 available sky fibers (red), the performance is limited by the number of 
sky fibers, therefore decreases as $\sqrt{N^{-1}+92^{-1}}$.  For a Coma 
nodded exposure, there are 167 sky fibers, therefore the $1\sigma$ limiting 
surface brightness (blue) follows the curve of $\sqrt{N^{-1}+167^{-1}}$.  
Stacking across different exposures is more complicated, and we estimate this 
performance by stacking across all 16 Coma nod exposures (green).  
We further smoothed the stacked nod exposures in the wavelength direction by 
a factor of 3 to mimic the $1\sigma$ limiting surface brightness we are able 
to achieve by 54 science exposures, which have $\sim3\times$ exposures as the 
nod exposures.  With the same number of science fibers 
as we used in our ICL targets, the $1\sigma$ limiting surface is 
about 28~\sb.  Finally, we estimate the $1\sigma$ limiting surface brightness 
achieved by stacking all ICL1 fibers in all 54 exposures by first 
subtracting off a high smoothed continuum model in each spectrum.   
The $1\sigma$ limiting surface brightness derived from ICL1, 27.9~\sb, 
is a close estimate of the real performance, and is roughly consistent 
with the predictions from the nodding exposures.  Since the surface 
brightness of three ICL regions is $\mu_g\approx25.3-26.2$, they are 
about at the same level of $10\sigma$ limiting surface brightness.  
In summary, we achieve a $1\sigma$ limiting surface brightness 
of 27.9~\sb~by spatially stacking the ICL spectra over 54 science exposures 
for our $\mu_g\approx25.3-26.2$~\sb~targets under the 
$\mu_g\approx22$~\sb~sky background.

\section{Stellar Population Modeling}

\subsection{Absorption Line Fitter}
 Our main tool for modeling spectra of galaxies and ICL in our sample 
 is the absorption line fitter \citep[{\tt alf},][]{Conroy2012, Conroy2014, 
 Conroy2016, Conroy2018}.   {\tt alf} enables stellar population modeling of the full 
 spectrum for stellar ages $>1$Gyr and for metallicities from $\sim-2.0$ to $+0.25$.  
 With {\tt alf} we explore the parameter space using a Markov Chain Monte Carlo 
 algorithm \citep[{\tt emcee},][]{ForemanMackey2013}.  The program now adopts 
 the MIST stellar isochrones \citep{Choi2016} and utilizes a new spectral library 
 that includes continuous wavelength coverage from $0.35-2.4\mu m$ over a wide 
 range in metallicity.  This new library, described in \citet{Villaume2017}, 
 is the result of obtaining new IRTF NIR spectra for stars in the MILES 
 optical spectral library \citep{SanchezBlazquez2006}.  Finally, theoretical 
 response functions, which tabulate the effect on the spectrum of 
 enhancing each of 18 individual elements, were computed using the 
 ATLAS and SYNTHE programs \citep{Kurucz1970, Kurucz1993}.  Further details of 
 these updates to {\tt alf} are described in \citet{Conroy2018}.
 With {\tt alf} we are able to fit a two burst star formation history, 
 the redshift, velocity dispersion, overall metallicity ([Z/H]), 18 
 individual element abundances, several IMF parameters, and a variety 
 of ``nuisance'' parameters.
 
\begin{figure}[t!]
\centering
\includegraphics[height=0.86\textheight]{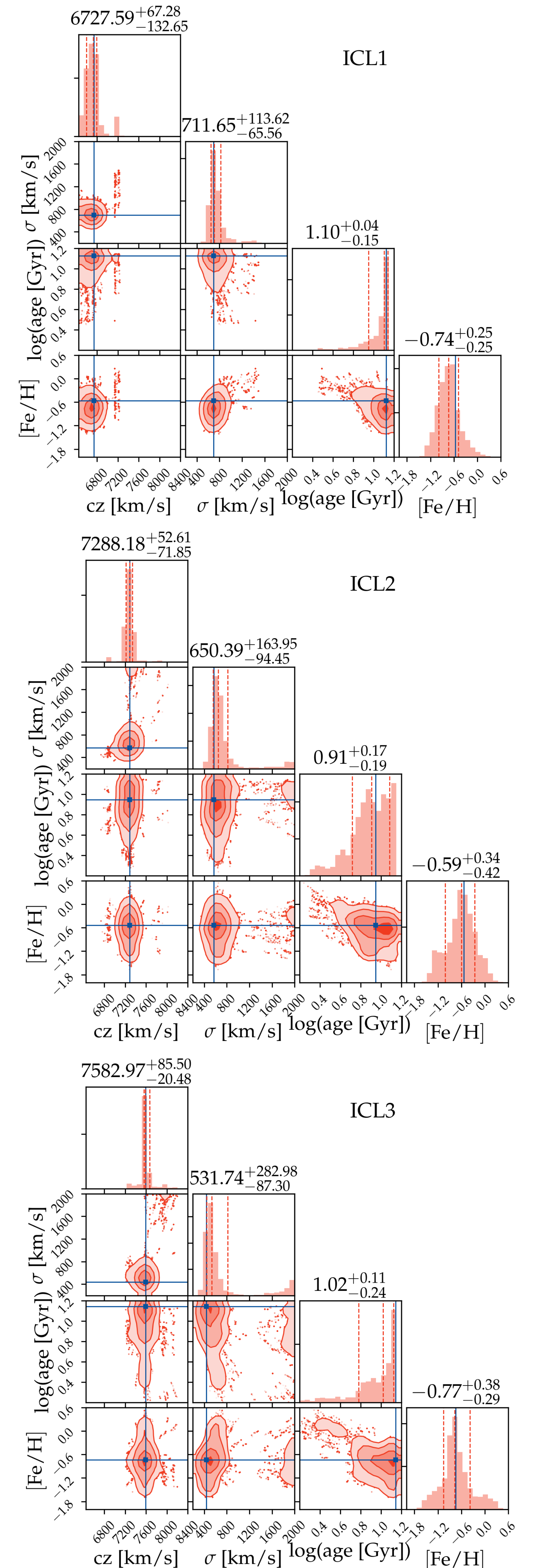}
\caption{
Projections of the posterior of recession velocity, velocity dispersion, 
log(age), and [Fe/H] from {\tt alf} in 1D and 2D histograms for ICL1, 
ICL2, and ICL3. Dashed lines and contours show the 16th,
50th and 84th percentiles of posteriors.  Blue lines represent 
the best fit parameters at $\chi^2_{min}$, which are used to 
generate best-fit model spectra.
}
\label{figure7}
\end{figure}

 Throughout this paper, we use {\tt alf} in a simplified mode.  
 Not all the parameters are included, but only the recession velocity, 
 age, overall metallicity [Z/H] and abundances of ~Fe, ~C, ~N, ~O, ~Mg, 
 ~Si, ~Ca, ~Ti, and Na. The IMF is fixed to the \citet{Kroupa2001} form.  
 Instead of adopting a two burst star formation history in the standard 
 model, the simplified mode adopts only a single age component.
 We adopt this approach due to the limited S/N of the data.  
 We adopt flat priors from $500-10500$~\kms~for recession velocity, 
 $10-2000$~\kms~for velocity dispersion, $1.0-14.0$~Gyr for age and 
 $-1.8-+0.3$ for [Fe/H].  The priors are zero outside these ranges. 
 For each spectrum, we normalize the continuum by fitting the ratio 
 between model and data in the form of a polynomial.  
 For spectra from IFU bundles that 
 are located within 50~kpc from the centers of BCGs, we use 
 the polynomial with order of $(\lambda_{max}$--$\lambda_{min})/100$\AA.  
 For ICL targets, due to the large velocity dispersion in the stellar 
 content, we adopt a 6th-order polynomial (see Appendix). 
 For each likelihood evaluation, the polynomial-divided 
 input spectra are matched with the model. 
 normalization occurs in two separate wavelength intervals, 
 $3800-4700$\AA~ and $4700-5600$\AA.  
 Pixels near bright sky lines in the blue were masked prior to the fitting. 

 Figure~5 shows model spectra for old and low metallicity stellar 
 populations with the same stellar age of 10~Gyr and [Fe/H] of $-0.7$, but  
 different velocity dispersions: 200, 400, and 600~\kms.  For a 
 typical ETG, the velocity dispersion in the central region is 
 roughly $200-300$~\kms.  The velocity dispersion in the ICL regions 
 are $>400$~\kms.  Compared to typical ETGs, the stellar 
 absorption features in regions with $>400$~\kms~become much shallower 
 with increasing velocity dispersion.  Therefore ICL spectra are noticeably 
 different than typical ETG spectra.

\begin{figure*}[t]
\centering 
\includegraphics[width=18cm]{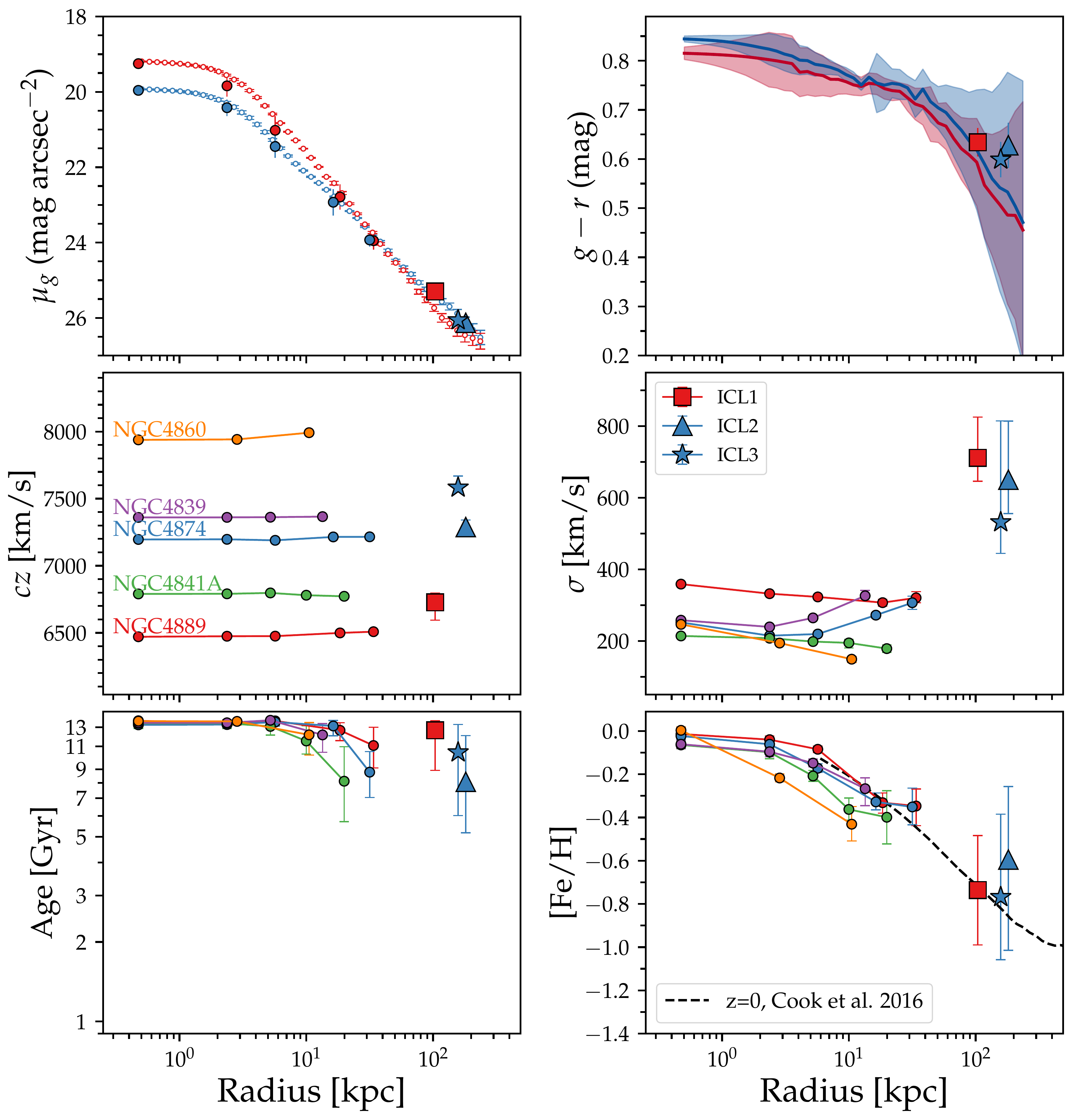}
\caption{
Top panels: $g$-band surface brightness profiles and $g-r$ color 
profiles for NGC~4889 (red) and NGC~4874 (blue), extended into the 
ICL regime; Middle panels: Best-fit recessional velocity $cz$ and 
velocity dispersion $\sigma$ as a function of radius;  
Bottom panels: Best-fit stellar age and [Fe/H] as a function of 
radius. Error bars enclose 16th and 84th percentile of the posteriors. 
Results for ICL regions are marked with the same color as that 
for their closest BCGs.  In the bottom right panel, the dashed line 
represents the shifted mean metallicity [Z/H] profile at $z=0$ in $\log\mathrm{(M_*/M_{\odot})\in[11.5,12.0)}$ from the Illustris 
simulations \citep{Cook2016}.  The profile is shifted down by 
$0.15$~dex for reasons explained in the text.
}
\label{sbp}
\end{figure*}

\begin{figure*}[t]
\centering
\includegraphics[width=16cm]{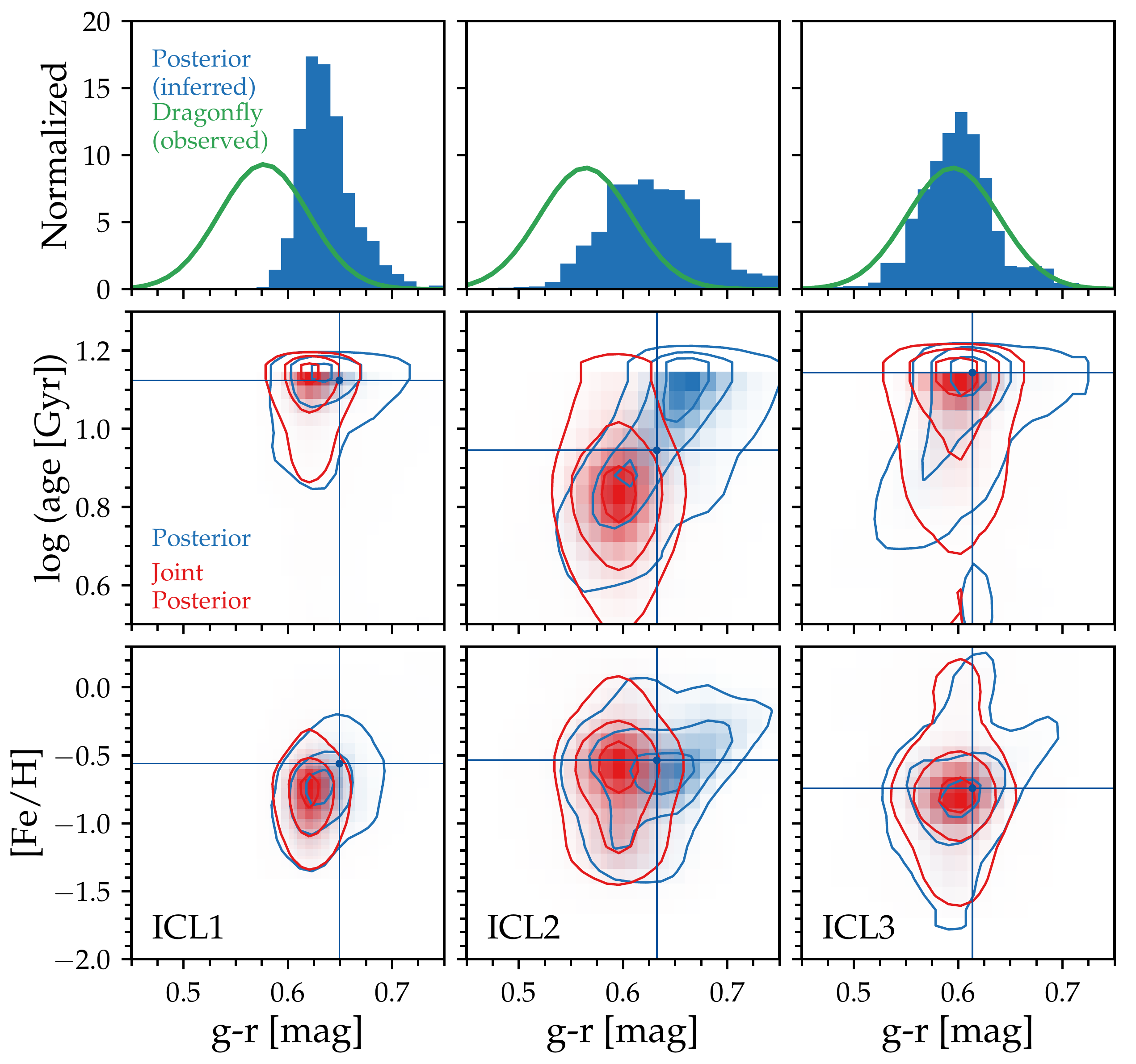}
\caption{
Top panel: 1D histogram of inferred $g-r$ color posterior from {\tt alf} 
(blue) for ICL1 (left), ICL2 (middle), and ICL3 (right), compared with 
the $g-r$ color from broadband images taken by the Dragonfly Telephoto 
Array (green).  Middle and Bottom panels: Projections of the posterior of log(age), [Fe/H] 
and $g-r$ in 2D histograms. Blue lines show the best fit parameters at 
$\chi^2_{min}$.  The joint posterior distributions from combining 
broadband colors and model spectra colors are shown in the same panels in red.
}
\label{poscolor}
\end{figure*}
\noindent

\begin{figure*}[t]
\centering
\includegraphics[width=17cm]{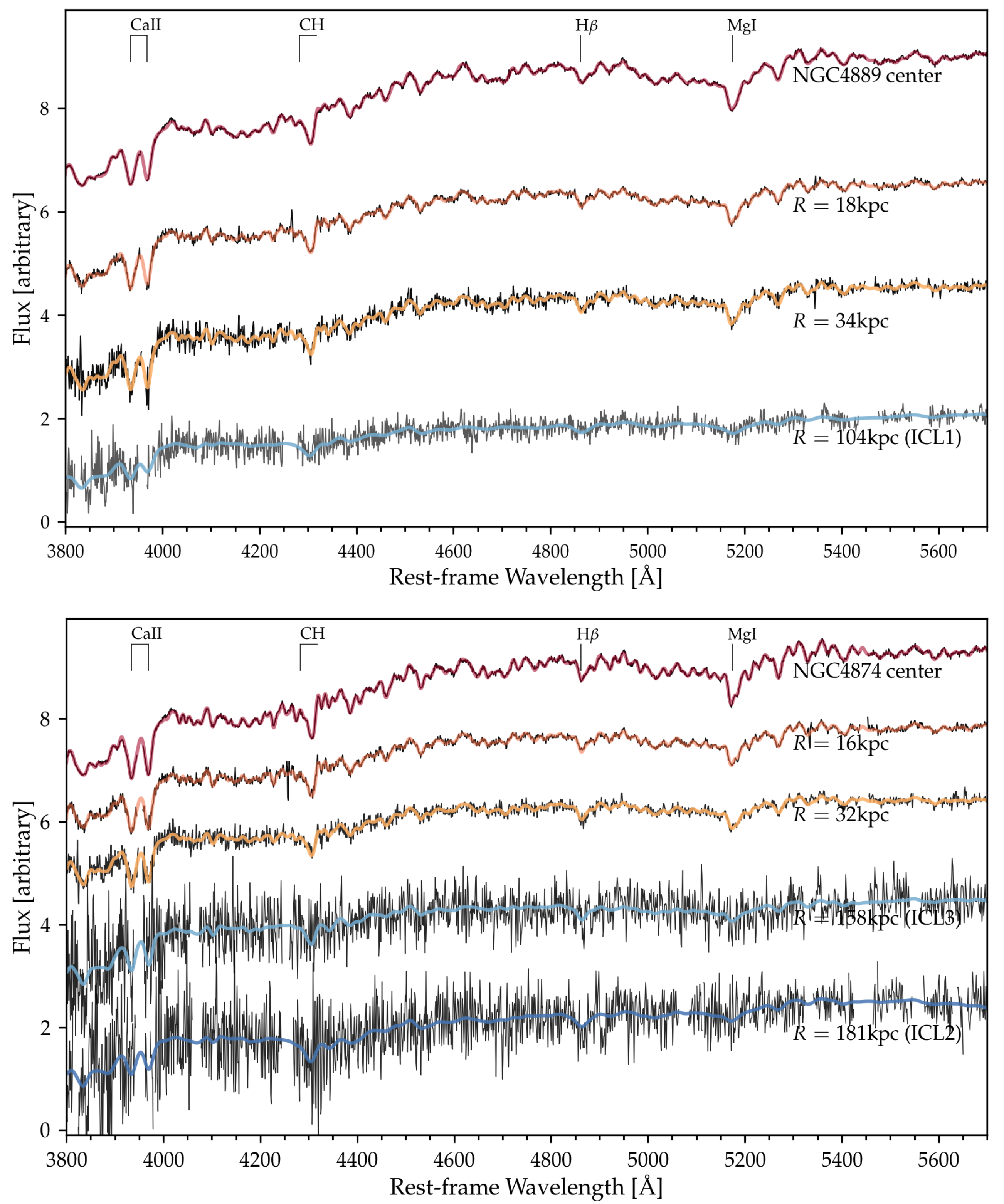}
\caption{
 Normalized median stacked spectra (black) at different distances to the centers of NGC~4889 
 (top panel) and NGC~4874 (bottom).  Strong absorption features are labeled. 
 Best-fit model spectra with parameters at minimum $\chi^2$ are shown in colors.   
}
\label{compspec}
\end{figure*}

\section{Results}
\subsection{Stellar Population of the ICL}

We now present our results from full spectra modeling.  The total on-source 
exposure time for each ICL regions is 13.5 hours.  The mean S/N ratio 
we achieved for ICL1, ICL2 and ICL3 in wavelength range from 
$4500-5000$\AA~are 21.6\AA$^{-1}$, 9.6\AA$^{-1}$, and 11.6\AA$^{-1}$.  
Figure~6 shows the median stacked spectra for ICL1, ICL2, and ICL3 
from the 54 15-minute 
exposures observed by the six Coma plates (Plate 8479, 8480, 8953, 9051).  
The spectrum shown in the 
figure is smoothed by a 3 pixel boxcar kernel.  Visually prominent 
stellar absorption features for old stellar populations have been captured 
by the best-fit model spectra (red), including Ca II H\&K lines, CH, 
H$\beta$, and MgI.  The red spectrum shows the best fit spectra by 
{\tt alf}.  In Figure~7 we show the projections of posteriors \citep{corner} 
for four parameters: recession velocity (cz), velocity dispersion, 
log(age) and [Fe/H].  The posterior distributions are well approximated 
by a Gaussian.
Dashed lines show the values of parameters at $16$th, 
$50$th and $84$th percentiles of posteriors.  Blue lines mark the values 
of parameters at minimum $\chi^2$.  Outliers are shown as dots.  

The Coma cluster has a median redshift of $cz=7090$\kms~\citep{Geller1999}.  
The recession velocities of ICL1, ICL2, and ICL3 are 
$6728^{+67}_{-133}$\kms, 
$7288^{+53}_{-72}$\kms, and $7583^{+86}_{-20}$\kms, respectively.
The derived velocity dispersions confirm that the stellar content in these 
regions belong to the Coma cluster.  All of the three ICL regions have high 
velocity dispersion: $712^{+114}_{-66}$\kms, $650^{+164}_{-94}$\kms, $532^{+283}_{-87}$\kms.
The ages of 
ICL1, ICL2, and ICL3 are $12.7^{+1.1}_{-3.7}$~Gyr, $8.1^{+4.0}_{-2.9}$~Gyr, 
and $10.4^{+2.9}_{-4.4}$~Gyr, respectively.  The iron abundances, [Fe/H], are 
$-0.74^{+0.25}_{-0.25}$, $-0.59^{+0.34}_{-0.42}$, and $-0.77^{+0.38}_{-0.29}$, 
respectively.  The stellar content in all three ICL regions is old and metal--poor.
We note that both the low metallicity and large velocity dispersion lead to less 
significant absorption features, making it more difficult to extract stellar 
population properties (see appendix fore more detail).

\subsection{Surface Brightness and Color Profiles}

\begin{figure*}[t]
\centering
\includegraphics[width=16cm]{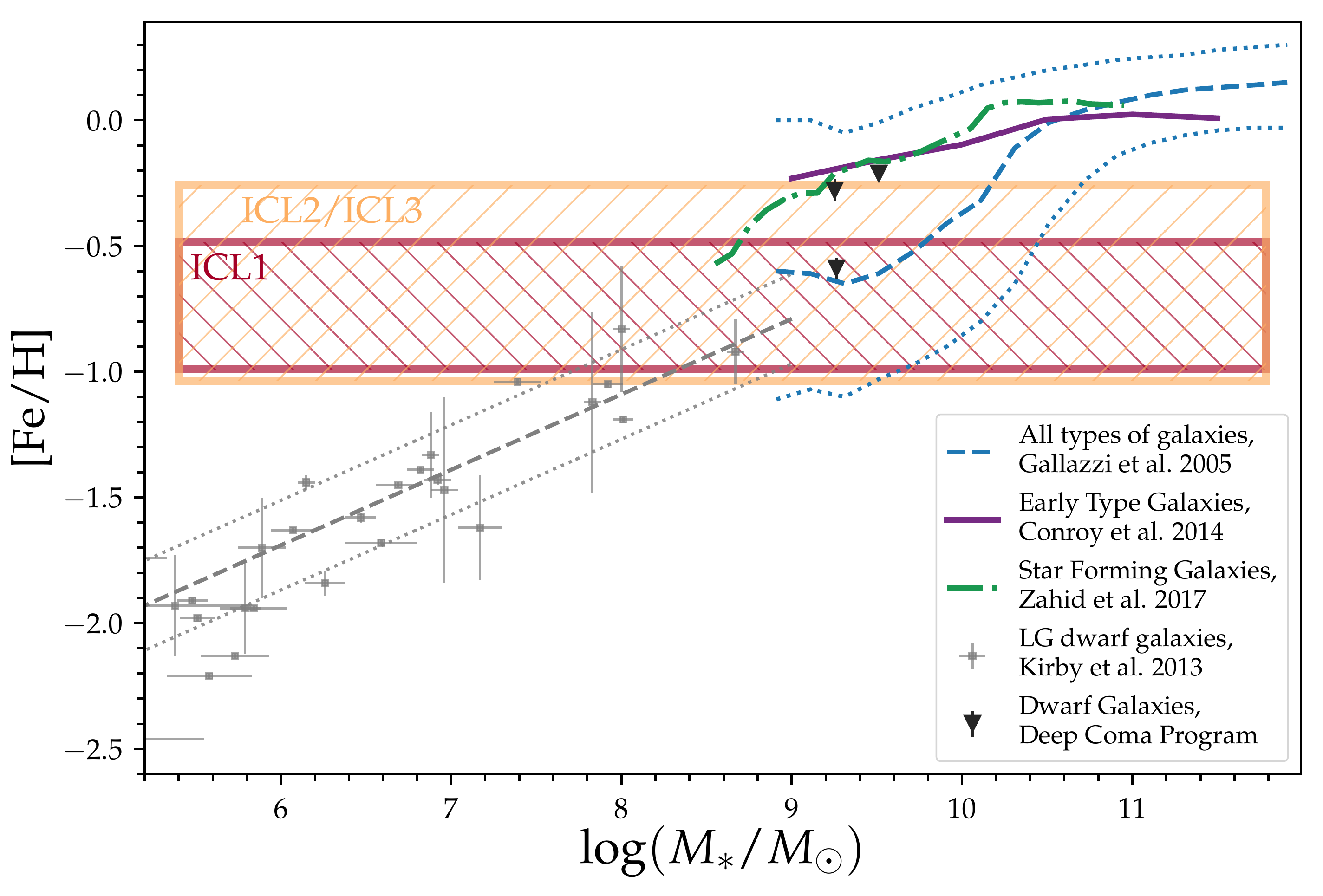}
\caption{
Relation between stellar mass and [Fe/H] for three dwarf elliptical galaxies in the 
Coma cluster (triangles), and previous results from the literature: gray symbols 
show Local Group dwarf galaxies from Kirby et al. (2013), and gray dashed and dotted 
lines represent the median,16th, and 84th percentiles of the metallicity distributions. 
Blue dashed and dotted lines show the median, 16th, and 84th percentiles of the 
metallicity distributions for various types of galaxies in Gallazzi et al. (2005). 
The purple line shows the stellar mass-–metallicity relation for early-type galaxies 
binned in stellar mass (Conroy et al. 2014).  The green line represents 
the relation for star-forming galaxies in SDSS \citep{Zahid2017}.
For the three ICL regions in the 
Coma cluster without stellar mass constraints, their median, 16th and 84th percentiles 
of [Fe/H] are shown as horizontal region. 
}
\label{figure11}
\end{figure*}

We derive the surface brightness profiles of NGC~4889 and NGC~4874 by 
performing the {\tt IRAF} \footnote{STSDAS is a product of the Space 
Telescope Science Institute, which is operated by AURA for NASA.} task 
{\tt ELLIPSE} \citep{Jedrzejewski1987} on the sky subtracted Dragonfly 
images.  Surrounding objects are aggressively masked iteratively using 
{\tt SExtractor} \citep{Bertin1996}.  The centers of galaxies are determined 
first, then the ellipticity and position angle are determined by the 
median values in radial range 10-100~kpc.  We extract the surface 
brightness profile along the major axis out to 300~kpc with fixed 
ellipticity and position angle, and correct them for Galactic 
extinction.  Error-bars on the surface brightness and $g-r$ color 
profiles are calculated by combining the uncertainty in intensity 
from the \texttt{ELLIPSE} procedure, and the intrinsic fluctuation 
of the background level obtained from aggressively masked Dragonfly 
images.  The $g$ band surface brightness and $g-r$ color profiles of 
NGC~4889 and NGC~4874 are shown in Figure~9.  $g$ band surface 
brightness predicted by the best-fit model spectra obtained by 
spatially stacking spectra in the first two plates, the 3rd and 4th 
plates, the 5th and 6th plates, and three ICL regions are shown as 
comparisons.  The results from two methods are consistent with each 
other within $1\sigma$.  Figure~9 also shows that the $g-r$ color 
of three ICL regions inferred from the best-fit {\tt alf} model are 
consistent with photometry from Dragonfly images.  Both BCG+ICL $g-r$ 
colors become bluer with increasing radius.  This is consistent with our 
spectroscopic results that the stellar population is more metal poor 
with increasing radius.

\subsection{Radial Variations}

In this section we compare the radial variations in the stellar population 
properties.  We first compare the median spatially stacked spectra among 
the central region of NGC~4889 and NGC~4874 (1st and 2nd plates), 
at around $0.5$--$1R_{\rm e}$ (3rd and 4th plates) and $1$--$2R_{\rm e}$ (5th and 6th plates), 
and the ICL regions in Figure~10.  Stacked spectra in Figure~6
are smoothed by a 3 pixel wide boxcar kernel, overplot by the best fit 
model spectra. The absorption feathers in the ICL spectra are visually 
shallower than the inner regions of BCGs, suggesting a higher velocity 
dispersion of stars in the ICL.  

The $g$ band image surface brightness profiles and $g-r$ color profiles 
derived from the Dragonfly images are shown in the top panels of Figure~\ref{sbp}.  
We then compare the stellar population properties as a function of radius 
in the lower four panels of Figure~\ref{sbp}, including their recession velocity 
($cz$), velocity dispersion ($\sigma$), stellar age and iron abundance 
[Fe/H].  We also include the radial trends of three massive ETGs in 
the Coma cluster observed in the Deep Coma program: 
NGC~4860, NGC~4841A, and NGC~4839.  The radial coverage 
of these three galaxies are $\sim0-30$~kpc.  The mean recession 
velocities of the three ICL regions are very close to their nearest 
BCGs, NGC~4889 and NGC~4874.  

For the velocity dispersion ($\sigma$), stellar age and iron abundance 
[Fe/H], a more clear view of the radial trends can be found in Figure~1.  
To study the spatial distribution of stellar population parameters from the
central regions of BCGs to the ICL, we make use of the MaNGA data cube 
that is rectified spatially in units of 0.5\asec~ spatial pixels 
(spaxels) for the observations of NGC~4889 and NGC~4874 in the first 
four plates, corresponding to the central regions of the two BCGs, 
$0.5R_{\rm e}$ of NGC~4874, and $1R_{\rm e}$ of NGC~4889. 
Spectra in adjacent spaxels are binned by Voronoi tessellation 
\citep{Cappellari2003}.  To control the bin size, the S/N achieved 
in each bin in the central regions of two BCGs are $210$\AA$^{-1}$ 
and $180$\AA$^{-1}$ for NGC~4889 and NGC~4874, respectively.  
The S/N achieved in each bin of the data cube on $1R_{\rm e}$ 
of NGC~4889 and $0.5R_{\rm e}$ of NGC~4874 are $120$\AA$^{-1}$ 
and $100$\AA$^{-1}$, respectively.  
For the $2R_{\rm e}$ of NGC~4889 and $1R_{\rm e}$ of 
NGC~4874, we use the ``mgCFrame'' files and spatially stacked all 
science fibers in the bundles, and achieve a S/N $\approx50$\AA$^{-1}$.
The 2D spatial distributions of the best-fit $\sigma$, age and [Fe/H] are 
displayed in Figure~1.  

Both BCGs+ICL structures 
have rising velocity dispersion profiles, suggesting that stars in the 
ICL trace the potential of the Coma cluster instead of any individual galaxy.
The stellar ages for the three massive galaxies and two BCG+ICL structures 
are generally old from the center to the outskirts.  This is consistent with 
previous observations of nearly flat stellar age profiles in ETGs 
\citep{Greene2015, Sanchez-Blazquez2007, Gu2018b}.  
The stellar age profiles show that both the in-situ and ex-situ components 
of the massive ETGs in Coma have old stellar populations.  The result 
highlights the effect of ``environmental quenching'' \citep{Peng2010}, 
and is consistent with the ``coordinated assembly'' picture in \citet{Gu2018b} 
that the massive ETGs in the central regions of galaxy clusters grow by 
accreting preferentially old stellar systems.  
In the bottom right panel of Figure~\ref{sbp}, 
We see declining [Fe/H] 
radial profiles for all of the three massive galaxies and both BCG+ICL 
structures. The [Fe/H] in the ICL regions are in general even more metal 
poor than the outskirts of BCGs at $1$--$2R_{\rm e}$.  
We compare the radial profile in the outskirts of 
two BCG+ICL structures in our sample with the prediction from 
the Illustris simulations \citep{Cook2016}.  The black dashed line in this 
panel shows the shifted mean projected profiles of [Z/H] in 
$z=0$ galaxies that are in the stellar mass bins of 
$\log\mathrm{(M_*/M_{\odot})\in[11.5,12.0)}$.  Note that we 
shift the mean profile since we are comparing only the 
gradients in the outskirts, and we need to account for 
the difference between the total metallicity in the simulation 
and the [Fe/H] in observations. The mean projected profiles of 
[Z/H] in the Illustris simulations is shifted down by 0.15~dex 
to match the [Fe/H] at 10~kpc.  This difference accounts for 
the difference between totally metallicity and [Fe/H]. 

\subsection{Combined Constraints from Spectra and Photometry}

The broadband $g-r$ color obtained from the Dragonfly Telephoto Array 
is used as an additional constraint to the stellar age and [Fe/H].  
We measure the color from the Dragonfly data within an aperture of 
$D=32$\asec, similar to the regions of our stacked spectra. The measured 
$g-r$ colors are corrected for Galactic extinction. They are 
$0.58\pm0.04$ mag, $0.56\pm0.04$ mag, and $0.60\pm0.04$ mag 
for ICL1, ICL2, and ICL3, respectively.  We assume 
the probability density of the observed $g-r$ colors to be a normal 
distribution and take the measured color and uncertainty as the mean 
and standard deviation.  The color distributions from Dragonfly 
photometry are shown in green in the top panels of Figure~\ref{poscolor}.  
Normalized 1D posterior distributions of the $g-r$ color derived 
from fitting the continuum-normalized spectra are shown in the top 
panels in blue. The differences are within $1\sigma$ photometric 
uncertainty.  We then re-weight the MCMC chains based on the 
probability density of the broadband $g-r$ color, and generate new 
posterior distributions by bootstrap resampling. The middle and 
bottom panels of Figure~\ref{poscolor} show the joint posterior distributions 
of log(age/Gyr) and [Fe/H] in red, respectively.  Slightly tighter 
constraints are achieved for ICL1 and ICL3. The jointly constrained 
stellar ages of ICL1, ICL2, and ICL3 are $12.4^{+1.2}_{-3.4}$~Gyr, 
$6.6^{+2.5}_{-2.4}$~Gyr, and $10.6^{+2.8}_{-4.4}$~Gyr, while the 
jointly constrained [Fe/H] are $-0.79^{+0.21}_{-0.25}$, 
$-0.64^{+0.30}_{-0.45}$, and $-0.79^{+0.32}_{-0.31}$, respectively.
Note that derived parameters only change in the case of the 
stellar age of ICL2.

\subsection{Dynamical Structure}
 
The Coma cluster is known to have a complex structure 
\citep{Mellier1988, Fitchett1987}.  Previous works found major 
substructures around the bright galaxy NGC~4839, indicating that 
there is continuing infall at the present day 
\citep{Fitchett1987,Colless1996,Neumann2001,Neumann2003}.
Furthermore, the core of the cluster comprises two giant cD galaxies, 
NGC~4889 and NGC~4874, and X–ray observations \citep{Adami2005} 
suggest that these galaxies are likely associated with two distinct 
substructures.  The line-of-sight velocity distributions of 
intracluster planetary nebulae in the central region of Coma 
\citep{Gerhard2007} also suggest an ongoing merger of distinct 
substructures associated with NGC~4889 and NGC~4874. Our results 
contribute to this picture with direct kinematic evidence that 
the core of Coma is, in fact, a double cluster with two distinct 
virialized components. The high velocity dispersions in the 
three ICL regions indicate that the stars in the ICL trace the 
gravitational potential of subclusters, not galaxies. However, 
the recession velocities are consistent with those of the two 
BCGs, with the difference in velocity between ICL1 and ICL2/ICL3 
are $\Delta cz\approx560-850$~\kms.  
This shows that stars in ICL1 are tracing a massive dark matter 
halo around NGC~4889, while stars in ICL2 and ICL3 are tracing a 
distinct halo around NGC~4874. Our results confirm that NGC~4889 
and NGC~4874 not only originate in separate clusters, but that 
those clusters are still distinct at the present day. 

\section{Discussion}

The stellar population in the outskirts of galaxies and the ICL contains 
important clues related to galaxy accretion and dynamical evolution. 
Simulations predict that ex-situ mergers are a major contributor to the 
diffuse light component, and its contribution to the outskirts and ICL 
regions are more significant for more massive galaxies 
\citep{Cooper2013,Lackner2012,Qu2017}.  For example, the IllustrisTNG 
simulations \citep{Pillepich2018} predict that a large fraction of 
stars ($>50\%$) in the outskirts of massive ETGs ($>30$~kpc) are 
formed ex-situ.  This fraction is even larger ($\sim90\%$) for stars 
beyond 100~kpc.  The radial profile of surface brightness and 
metallicity of galaxies provide us important clues to their accretion 
histories.  For example, tidally stripped stars at large radii are 
expected to flatten metallicity and surface brightness profiles 
\citep{DiMatteo2009,Vogelsberger2014,Genel2014,Cook2016}.
\citet{Cook2016} shows that in the Illustris Simulations, the 
radial profiles of metallicity and surface brightness flatten 
from $z=1$ to the current epoch due to the accretion of stars 
into the outskirts of galaxies.  We compare radial 
trends of the Coma BCG+ICL structures with the simulations.  
Despite the large uncertainties of metallicity in the ICL regions, the 
radial of [Fe/H] trends at 100-200~kpc is consistent 
with the prediction from simulations.  

If the light at $100-200$~kpc is indeed dominated by ex-situ 
components as predicted by the simulations 
\citep[e.g.][]{Pillepich2018}, stellar populations in the 
ICL regions can shed light on the progenitors of ICL. 
ETGs experienced complex assembly histories, however, they 
obey tight scaling relations, including the relation between 
the stellar mass and stellar metallicity, which provides important 
clues to their star formation and chemical enrichment history 
\citep{Lu2017, Ma2016, Kirby2013}.  By comparing the [Fe/H] of 
ICL we can infer their possible progenitors.  Figure~11 shows 
the relation between stellar mass and stellar metallicity (MZR). 
In Figure~11, we draw a horizontal region across nearly all 
stellar mass with the vertical range covering the 18th and 
84th percentiles of[Fe/H] in three ICL regions.  The [Fe/H] of 
the ICL is compared to other populations, including three 
dwarf elliptical galaxies in the Coma cluster in our sample, 
ETGs stacked in stellar mass bins \citep{Conroy2014}, star-forming galaxies in SDSS 
\citep{Zahid2017}, and the MZR from \citep{Gallazzi2005}, which 
covers both star-forming and quiescent SDSS galaxies, and dwarf 
galaxies in the Local Group \citep{Kirby2013}. Data points from 
\citet{Gallazzi2005} and \citet{Zahid2017} represent the total 
metallicity instead of iron abundances.  Note that for the more 
massive galaxies, the figure shows the relation between their 
stellar mass and [Fe/H] or total metallicity in the inner region.

The [Fe/H] of ICL regions are similar to the [Fe/H] of three 
dwarf elliptical galaxies in the Coma cluster.  If we estimate 
the stellar mass of the ICL progenitors based on the MZR, and assume 
the progenitors have early truncated star formation histories as 
indicated by the stellar age of three ICL regions, the stellar 
mass of possible progenitors cover a wide range from 
$\sim10^8M_{\odot}$ to $3\times10^9M_{\odot}$.  
The total ICL stellar mass is $\approx10^{12}M_{\odot}$, 
therefore the BCGs would need to merger with $3\times10^3-10^4$ such 
dwarf galaxies if they are the only contributors.  Considering we are 
comparing with the [Fe/H] in the galaxy centers, and their outskirts in general have lower [Fe/H], 
it is likely that the stars of the ICL come partly from the outskirts of more massive galaxies.  
Therefore, the build up of ICL could be from partial tidal stripping of massive galaxies, and/or 
the disruption of dwarf galaxies with stellar mass above $\sim10^8M_{\odot}$.  But it is unlikely 
that the ICL form directly from major mergers of massive galaxies. 

The ICL1 spectrum is located between NGC~4889 and NGC~4874.  Using a spectral model 
with a single velocity component, we cannot rule out the possibility that the stellar 
contents in ICL1 consist of two velocity components, both with lower velocity 
dispersion.  With higher S/N data, we should be able to fit the spectra with a two 
component model to confirm our conclusion.  

\section{Summary}

We have presented the stellar population analysis through full optical spectral 
modeling for three ICL regions in the Coma cluster that are located between 
$100$--$200$~kpc from their nearest BCGs.  We have measured their recession velocities, 
velocity dispersion, stellar ages, and iron abundances using spectra obtained as 
part of the Deep Coma Program within the SDSS-IV/MaNGA survey.  
Based on their line-of-sight velocity and velocity dispersion, 
the three ICL regions are associated with two distinct subclusters centered on 
NGC~4889 and NGC~4874. 
For the BCG+ICL structures, the radial profiles of 
stellar age are old and flat, and the radial profiles of [Fe/H] decline with increasing 
radius.  The stellar populations in the ICL regions are all old and metal-poor. The 
[Fe/H] of three ICL regions are slightly more metal poor compared to the outskirts 
(10--30~kpc) of massive ETGs in the Coma cluster. From the derived stellar age and 
metallicity, the build-up of ICL is likely to be through the accretion of low mass 
galaxies that have ended their star formation early on, or partial tidal stripping 
of massive galaxies, instead of major mergers of massive galaxies.

\acknowledgements
M.G. acknowledges support from the National Science Foundation 
Graduate Research Fellowship.  
C.C. acknowledges support from NASA grant NNX15AK14G, NSF grant
AST-1313280, and the Packard Foundation. 
MAB. acknowledges NSF Award AST-1517006.
The computations in this paper were run on the Odyssey cluster supported 
by the FAS Division of Science, Research Computing Group at Harvard University.

Funding for the Sloan Digital Sky Survey IV has been provided by the 
Alfred P. Sloan Foundation, the U.S. Department of Energy Office of
Science, and the Participating Institutions. SDSS-IV acknowledges 
support and resources from the Center for High-Performance Computing 
at the University of Utah. The SDSS website is www.sdss.org.

SDSS-IV is managed by the Astrophysical Research Consortium for the 
Participating Institutions of the SDSS Collaboration including the 
Brazilian Participation Group, the Carnegie Institution for Science, 
Carnegie Mellon University, the Chilean Participation Group, the 
French Participation Group, Harvard-Smithsonian Center for Astrophysics, 
Instituto de Astrof\'isica de Canarias, The Johns Hopkins University, 
Kavli Institute for the Physics and Mathematics of the Universe 
(IPMU) / University of Tokyo, Lawrence Berkeley National Laboratory, 
Leibniz Institut f\"ur Astrophysik Potsdam (AIP),  
Max-Planck-Institut f\"ur Astronomie (MPIA Heidelberg), 
Max-Planck-Institut f\"ur Astrophysik (MPA Garching), 
Max-Planck-Institut f\"ur Extraterrestrische Physik (MPE), 
National Astronomical Observatories of China, New Mexico State University, 
New York University, University of Notre Dame, 
Observat\'ario Nacional / MCTI, The Ohio State University, 
Pennsylvania State University, Shanghai Astronomical Observatory, 
United Kingdom Participation Group, Universidad Nacional 
Aut\'onoma de M\'exico, University of Arizona, 
University of Colorado Boulder, University of Oxford, 
University of Portsmouth, University of Utah, University of Virginia, 
University of Washington, University of Wisconsin, 
Vanderbilt University, and Yale University.


\appendix
In this Appendix we examine our ability to recover the recessional 
velocity, velocity dispersion, age and overall metallicity for ICL-like 
spectra with {\tt alf}. The ICL presents a few related unique challenges 
due to the low S/N and very high velocity dispersions.  

We begin by 
exploring the effect of the continuum normalization on the derived 
parameters.  Our default approach is to set the degree of polynomial as $(\lambda_{max}$--$\lambda_{min})/100$\AA, which means a 10th order 
polynomial in the wavelength range $3800-5700$\AA.  We test the 
sensitivity of the results to the polynomial degree in Figure~12.  We have 
constructed a mock spectra dataset with ten realizations at a range of 
S/N from 10 to 100\AA$^{-1}$.  We assumed the mock spectra have a true 
recessional velocity of 7200~\kms, a velocity dispersion of 800~\kms, 
an age of 10 Gyr and a metallicity of [Z/H]$=-0.5$. The data are fit 
over two wavelength ranges,  4000--4700\AA~ and 4700--5700\AA~.  
The results are shown in Figure~12 as a function of the highest degree of 
the polynomial used in fitting the continuum.  To test the appropriate 
order of polynomial term to use, we fit mock spectra with different 
assumption of this order from three to ten.  It terms 
out this degree is crucial to our results.  The velocity dispersion and low 
metallicity make it difficult to extract absorption features.  If a high order 
polynomial is used to fit the continuum, real absorption features could be 
misinterpreted as part of the continuum, thus producing a biased result.  
As shown in Figure~12, a high 
order polynomial slightly biases age and metallicity, especially for 
low S/N spectra, and results in larger errorbars.  
The figure shows that the recessional velocity and velocity 
dispersion can be recovered at good precision for mock spectra at all S/N.  
The biased recessional velocity from the true value is within 50~\kms, and 
the biased velocity dispersion are no more than 20~\kms~
different if we adopt 
a highest degree of polynomial smaller than 6.  Even for S/N$\leq20$ mock spectra, 
age and [Z/H] can be recovered well if we choose the right degree of polynomial 
to fit the continuum. In this paper, we choose to use a polynomial with the 
highest degree of 6 to fit the ICL spectra.  The mean bias on age is 0.6 and 
0.3~Gyr at S/N$=10$\AA$^{-1}$ and S/N$=20$\AA$^{-1}$, and the mean bias on 
[Z/H] is 0.04 and 0.01~dex, respectively. 

\begin{figure}[t!]
  \centering
  \includegraphics[width=12cm]{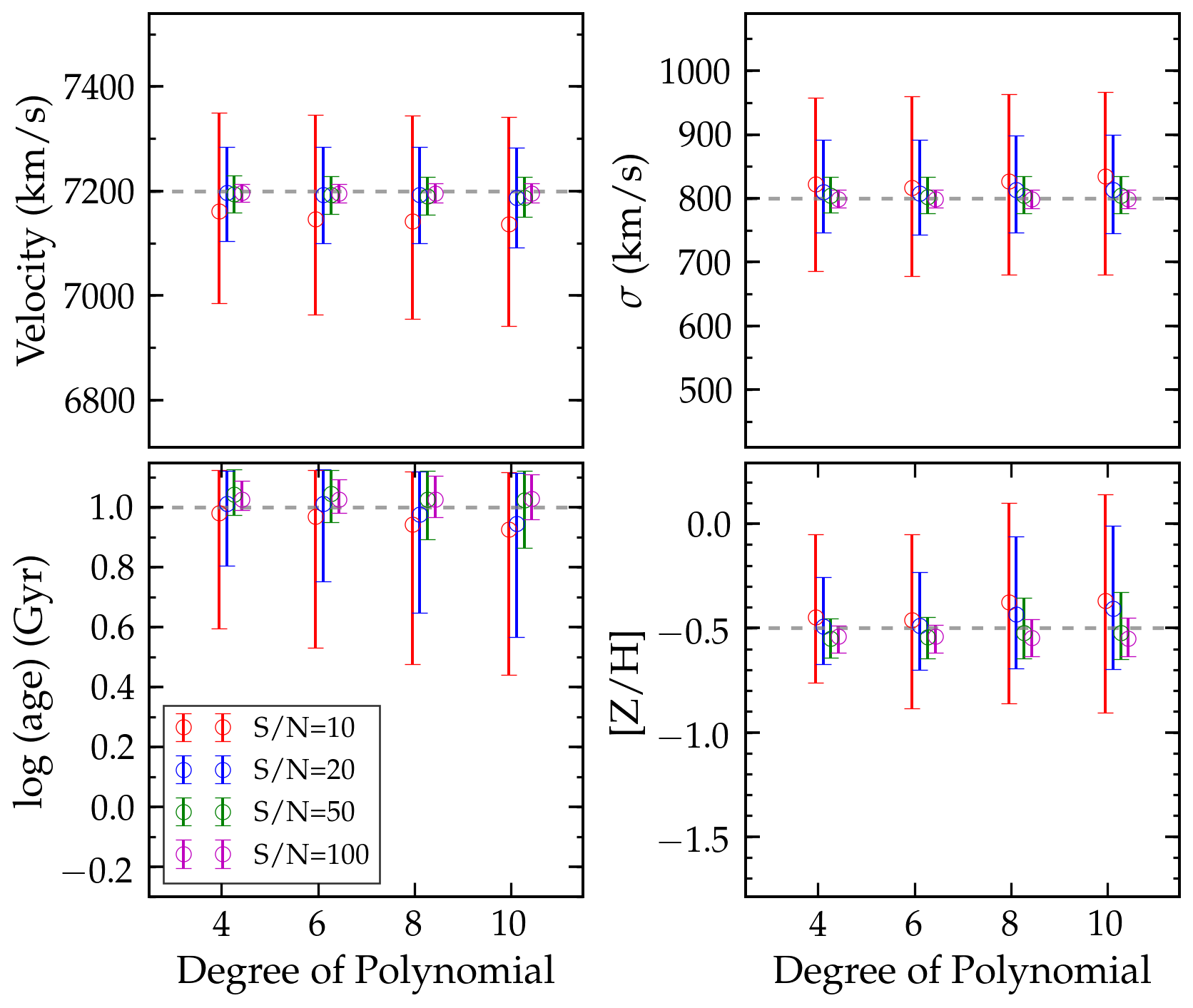}
  \caption{
  Test of recovery of recessional velocity, velocity dispersion, age and 
  overall metallicity [Z/H] with high velocity dispersion mock spectra 
  by tuning the order of polynomial used to fit the continuum. 
  The default degree of polynomial is $(\lambda_{max}$--$\lambda_{min})/100$\AA.   
  We constructed 10 realizations at each S/N.  The figure shows the mean values 
  of these 10 realizations.  Red, blue, green and purple symbols show parameters 
  at $50$~th percentile with spectra at S/N of 10, 20, 50 and 100\AA$^{-1}$, 
  respectively. Error bars enclose 16~th to 84~th percentiles.  Gray dashed lines 
  indicate the input values.  For ICL spectra we use a $6$~th order polynomial.
  }
  \label{figa1}
\end{figure}

\begin{figure}[t!]
  \centering
  \includegraphics[width=18cm]{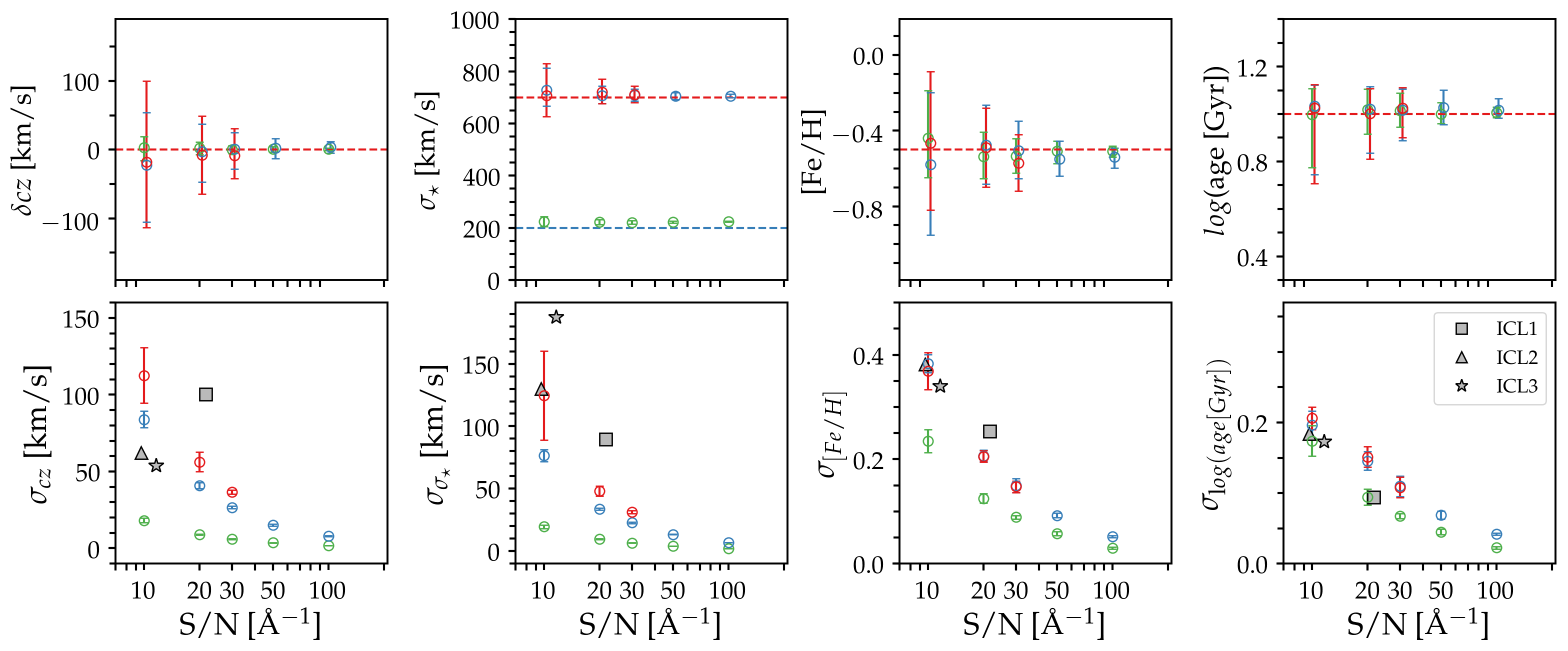}
  \caption{
  Recovery of parameters using mock spectra of systems with 
  low (200\kms, green) and high (700\kms, red and blue) velocity 
  dispersion as a function of S/N. We constructed 10 realizations 
  at each S/N. 
  Top panels: recovered velocity, velocity dispersion, [Fe/H], 
  and stellar age, compared to the input values (dashed lines).  
  Error bars enclose 16~th to 84~th percentiles; Bottom Panels: 
  typical uncertainty on the recovered parameters.  
  For stellar populations with high velocity dispersion (700\kms), 
  their recovered parameters have larger errorbars.  
  The uncertainty of stellar population parameters from fitting the ICL 
  spectra are compared (gray).
}
  \label{figa2}
\end{figure}

The stellar absorption features in the ICL spectra are much weaker 
compared to typical ETG spectra due to the higher velocity dispersion 
of the former (Figure~5).   As a consequence, the uncertainties on 
derived parameters are likely to be higher for the ICL than for ETGs, 
at a given S/N.  We explore this issue in Figure~13, where we show 
recovered parameters, and uncertainties, as a function of S/N for an 
ETG-like dispersion (200\kms) and an ICL-like dispersion (700\kms).  
We also include in the lower panels the actual uncertainties and S/N 
values for the three ICL regions in the Coma cluster.
t S/N of $10-20$\AA$^{-1}$, which 
is similar to the S/N we achieved in the spatially stacked ICL spectra, the 
uncertainty in [Fe/H] is about 0.1~dex larger than that derived from fitting 
typical ETG spectra.  
We also construct mock spectra with high (700~\kms) velocity dispersion, 
but the S/N in 3800\AA~to 4000\AA~is only $60\%$ of that in the rest part 
of spectra (red), similar to the non-uniform S/N in the ICL data.  
Although the uncertainties on [Fe/H] and stellar age are similar to the 
case of uniform S/N, the error constraints on recession velocity and 
velocity dispersion are worse.  
The results of three ICL regions are shown as a comparison.   
Their error constraints are closer to what we found in the mock data 
with high velocity dispersion and nonuniform spectra uncertainty. 
\end{document}